\documentclass[twocolumn,english,aps,prb,twocolum,superscriptaddress,natbib,bibnotes,amsmath,amssymb,floatfix,groupedaddress,footinbib]{revtex4-2}

\usepackage[markup=nocolor, authormarkupposition=left]{changes} 
\usepackage{soul}
\usepackage[utf8]{inputenc}
\usepackage[english]{babel}
\usepackage{amsmath,amsfonts,amssymb}
\usepackage[T1]{fontenc}
\usepackage{url}

\usepackage{amsmath}
\usepackage{siunitx}
\usepackage[version=4]{mhchem}
\usepackage{amsfonts}
\usepackage{amssymb}

\usepackage[colorlinks=true,citecolor=blue,linkcolor=magenta]{hyperref}

\usepackage{epstopdf}
\usepackage{graphicx}
\graphicspath{{./Figures/}}

\usepackage{changes}

\begin{document}

\title{Optical next generation reservoir computing}

\author{Hao Wang$^{1,2,\ast}$, 
        Jianqi Hu$^{1,4,\ast,\dagger}$, 
        YoonSeok Baek$^{1}$,       
        Kohei Tsuchiyama$^{1,3}$,
        Malo Joly$^{1}$,
        Qiang Liu$^{2,\dagger}$,  
        and Sylvain Gigan$^{1,\dagger}$}
\affiliation{
$^1$Laboratoire Kastler Brossel\char`,{} École Normale Supérieure - Paris Sciences et Lettres (PSL) Research University\char`,{} Sorbonne Université\char`,{} Centre National de la Recherche Scientifique (CNRS)\char`,{} UMR 8552\char`,{} Collège de France\char`,{} 24 rue Lhomond\char`,{} 75005 Paris\char`,{} France.\\
$^2$State Key Laboratory of Precision Space-time Information Sensing Technology\char`,{} Department of Precision Instrument\char`,{} Tsinghua University\char`,{} Beijing 100084\char`,{} China.\\
$^3$Department of Information Physics and Computing\char`,{} Graduate School of Information Science and Technology\char`,{} The University of Tokyo\char`,{} 7-3-1 Hongo\char`,{} Bunkyo-ku\char`,{} Tokyo 113-8656\char`,{} Japan.\\
$^4$Present address: Swiss Federal Institute of Technology Lausanne (EPFL)\char`,{} CH-1015 Lausanne\char`,{} Switzerland.\\
}

\maketitle

\noindent\textbf{\noindent
Artificial neural networks with internal dynamics exhibit remarkable capability in processing information. Reservoir computing (RC) is a canonical example that features rich computing expressivity and compatibility with physical implementations for enhanced efficiency. Recently, a new RC paradigm known as next generation reservoir computing (NGRC) further improves expressivity but compromises its physical openness, posing challenges for realizations in physical systems. Here we demonstrate optical NGRC with computations performed by light scattering through disordered media. In contrast to conventional optical RC implementations, we drive our optical reservoir directly with time-delayed inputs. Much like digital NGRC that relies on polynomial features of delayed inputs, our optical reservoir also implicitly generates these polynomial features for desired functionalities. By leveraging the domain knowledge of the reservoir inputs, we show that the optical NGRC not only predicts the short-term dynamics of the low-dimensional Lorenz63 and large-scale Kuramoto-Sivashinsky chaotic time series, but also replicates their long-term ergodic properties. Optical NGRC shows superiority in shorter training length, increased interpretability and fewer hyperparameters compared to conventional optical RC based on scattering media, while achieving better forecasting performance. Our optical NGRC framework may inspire the realization of NGRC in other physical RC systems, new applications beyond time-series processing, and the development of deep and parallel architectures broadly.
}

\section*{Introduction} 

\noindent{Dynamical} systems, which receive external stimuli and responsively react to them, possess remarkable capacity to manipulate and process information \cite{werbos1990backpropagation,mante2013context,dambre2012information,hu2023tackling}. As a nonlinear dynamical system, reservoir computing (RC) often builds upon recurrent neural networks (RNNs), utilizing the dynamics of internal reservoir states by a weighted summation to achieve desired functionalities \cite{6789852,jaeger2001echo,jaeger2004harnessing,Sussillo_Abbott_2009,yan2024emerging}. The concept of RC not only finds various applications in time series forecasting \cite{jaeger2004harnessing,pathak2018model,bollt2021explaining,pammi2023extreme}, classification \cite{bianchi2020reservoir}, prediction \cite{lu2017reservoir}, attractor manipulation \cite{kim2021teaching} and robots control \cite{antonelo2008event}, but also connects computing theory, machine learning, neuroscience, biology and physics broadly \cite{nakajima2021reservoir}.

What makes RC so appealing is, in part, its physical compatibility. A broad range of physical mechanisms and substrates have been harnessed to implement reservoirs (Fig. \ref{Fig1}a) \cite{tanaka2019recent}, including analog electronics \cite{appeltant2011information,du2017reservoir}, spintronic oscillators \cite{torrejon2017neuromorphic,grollier2020neuromorphic}, biological organoids \cite{cai2023brain}, and many more. All of these physical implementations aim at energy-efficient and high-throughput non-von Neumann architectures \cite{markovic2020physics,mehonic2022brain,jaeger2023toward,stern2023learning}. Among others, optical computing is of particular interest \cite{wetzstein2020inference,shastri2021photonics,gigan2022imaging}, which employs photons as information carrier and light-matter interactions as processors, thereby exploiting the parallelism, energy efficiency and fast dynamics of light \cite{mcmahon2023physics}. Within optical computing, optical RC has a history of exploration for over a decade \cite{vandoorne2008toward,paquot2012optoelectronic,larger2012photonic} and can be broadly classified into two types, i.e., delay-based reservoirs \cite{paquot2012optoelectronic,larger2012photonic,martinenghi2012photonic,brunner2013parallel,vinckier2015high,duport2016fully,larger2017high,penkovsky2019coupled} and spatial-distributed reservoirs \cite{vandoorne2008toward,vandoorne2014experimental,brunner2015reconfigurable,bueno2018reinforcement,antonik2019human,dong2019optical,rafayelyan2020large,sunada2021photonic}. The former relies on either a single \cite{paquot2012optoelectronic,larger2012photonic,martinenghi2012photonic,brunner2013parallel,vinckier2015high,duport2016fully,larger2017high} or multiple \cite{penkovsky2019coupled} nonlinear devices with time-delayed feedback to create virtual reservoir nodes in the time domain. The latter encompasses reservoir systems built on semiconductor optical amplifiers \cite{vandoorne2008toward}, integrated delay line networks \cite{vandoorne2014experimental}, diffractively coupled optical elements \cite{brunner2015reconfigurable}, spatial light modulators (SLM) and cameras \cite{bueno2018reinforcement,antonik2019human}, as well as multiple light scattering media \cite{dong2019optical,rafayelyan2020large,sunada2021photonic}.

Often, new propositions on RC algorithms also influence and guide the designs of physical reservoir computing. 
% Taken together, the reservoir's physical implementations arguably follow the lead in RC's algorithmic advancements. 
For instance, a recent proposal of graph reservoir computing \cite{gallicchio2020fast} has been implemented in a topology of analog random resistive memory cells, achieving orders of magnitude higher energy efficiency compared to its digital courterpart \cite{wang2023echo}. Another example is the realization of deep reservoir computing networks in optics \cite{lupo2023deep,shen2023deep}, where the multi-timescale dynamics of stacked layers yields better computing performance \cite{gallicchio2017deep}. Recently, a new RC paradigm, known as `next generation reservoir computing' (NGRC) \cite{gauthier2021next}, has been proposed, which defines a reservoir feature directly from the domain knowledge of the original data \cite{pyle2021domain}. True to its namesake, NGRC requires no more actual reservoirs for information mixing, but rather computes polynomial terms directly from the time-delayed inputs (Fig. \ref{Fig1}b). Such digital NGRC has been trained to outperform traditional RC in benchmark forecasting and prediction tasks, even with less training data and time \cite{pyle2021domain,gauthier2021next}. However, such a powerful architecture with growing prevalence in RC to date lacks physical realizations, partly due to the challenge of synthesizing these reservoir nodes explicitly.

In this work, we demonstrate an optical NGRC scheme based on light scattering through disordered media. Specifically, we drive our optical system with time-delayed inputs (Fig. \ref{Fig1}c), as opposed to feeding current inputs and reservoir states in almost all previous physical RC implementations. Instead of generating polynomial features directly as in digital NGRC, such a refinement also allows the optical setup to produce expanded polynomial features, embedded in the generic high-dimensional speckle intensity representations (Fig. \ref{Fig1}e). Optical NGRC features a multitude of advantages over conventional optical RC \cite{rafayelyan2020large} in processing time-series data. First, we demonstrate its efficacy in the short-term prediction of low-dimensional Lorenz63 and large-scale Kuramoto-Sivashinsky (KS) chaotic time series, achieving better prediction performance while using less than one tenth of the training data and a smaller reservoir compared to the previous state-of-the-art in optical RC \cite{rafayelyan2020large}. Moreover, the optical NGRC also replicates the `climate' of the original manifolds in the long-term dynamics, which acts as a photonic surrogate model. Furthermore, we show that the optical NGRC can accurately infer unmeasured state variables in observer prediction applications, outperforming standard digital interpolation methods. The optical NGRC demonstrated in this work delivers interpretable results by synthesizing features of time-delayed inputs through the optical system and then linearly combine them for versatile functionalities. Though our scheme is an indirect form of digital NGRC, it offers substantial compatibility with physical computing systems, thereby providing insights for tailoring various other physical reservoirs.

\section*{Results} 
\begin{figure*}[!htp]
  \centering{
  \includegraphics[width = 0.9\linewidth]{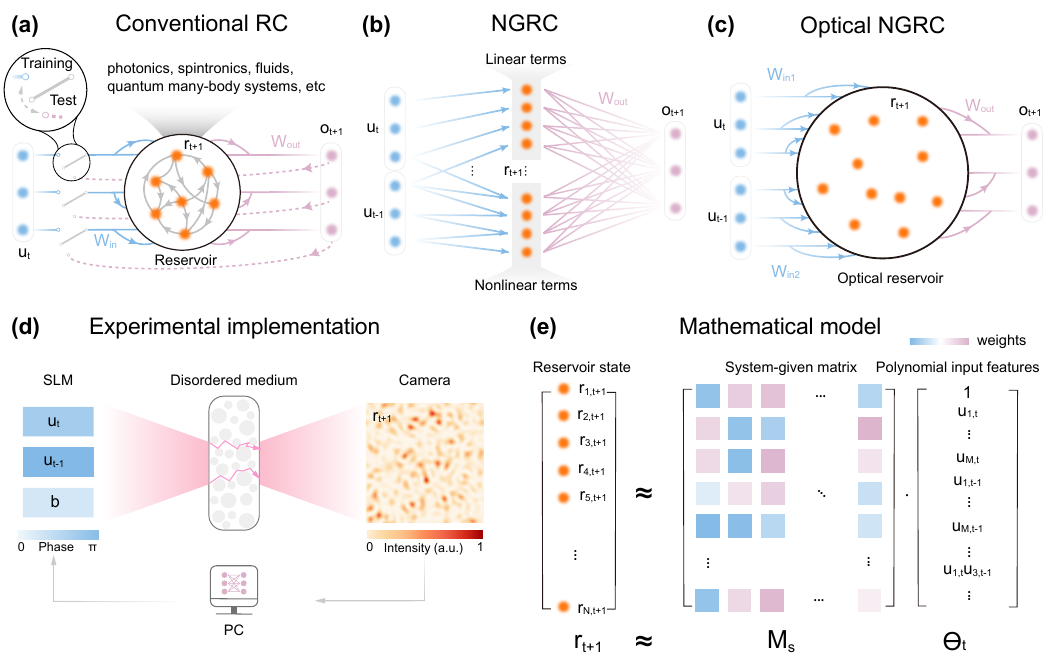}
  } 
    \caption{\noindent\textbf{Optical next generation reservoir computing.}  \textbf{(a)} Reservoir computing (RC) is a special type of recurrent neural network and is compatible with various physical implementations. In the training phase, RC sequentially maps the current input ($\boldsymbol{u}_t$, blue) and the current reservoir state ($\boldsymbol{r}_{t}$) into the next reservoir state ($\boldsymbol{r}_{t+1}$, orange). After that, only a linear readout layer $\boldsymbol{W}_{out}$ is trained to match $\boldsymbol{\hat{o}}_{t} = \boldsymbol{W}_{out}\boldsymbol{r}_{t}$ with the desired output ($\boldsymbol{o}_{t}$, purple), which is often the input (i.e. $\boldsymbol{o}_{t}=\boldsymbol{u}_{t}$) in the time series prediction tasks. In the prediction phase, by feeding back the predicted output to the input, RC can autonomously evolve as a dynamical system. \textbf{(b)} Different from the conventional RC scheme, next generation reservoir computing (NGRC) directly synthesizes reservoir features by constructing the polynomial functions of the time-delayed inputs (e.g., $\boldsymbol{u}_{t}$ and $\boldsymbol{u}_{t-1}$), without relying on an actual reservoir. \textbf{(c)} Similar to the digital NGRC, the proposed optical NGRC also drives the optical reservoir with time-delayed inputs. The optical process generates the polynomial features of the inputs implicitly. \textbf{(d)} The schematic experimental setup for the optical NGRC. First, input data at the current and the previous  time steps ($\boldsymbol{u}_t$ and $\boldsymbol{u}_{t-1}$) as well as a bias $\boldsymbol{b}$, are encoded onto the phase front of a laser beam via a spatial light modulator (SLM). Then, the modulated coherent light illuminates a disordered scattering medium, which provides rich mixing of the input and generates speckle patterns at the output. Finally, the reservoir features are obtained by measuring the intensity of the speckles in a camera. A computer (PC) is used to interface the SLM and the camera, as well as training and implementing readout layer.
    \textbf{(e)} The mathematical model of optical NGRC. The nonlinear, implicit reservoir speckle features $\boldsymbol{r}_{t+1}$ can be approximated as the multiplication of a system-given matrix $\boldsymbol{M}_s$ by a library of explicit polynomial feature terms $\boldsymbol{\Theta_t}$}.
    \label{Fig1}
\end{figure*}

\noindent\textbf{Principle.} We begin by briefly introducing the concept of RC, which is a RNN with fixed and random connectivity (Fig. \ref{Fig1}a). For input data $\boldsymbol{u}_t=(u_{1,t}, u_{2,t},...,u_{M,t}) \in \mathbb{R}^M$ and the internal reservoir states $\boldsymbol{r}_t=(r_{1,t},r_{2,t},...,r_{N,t})\in \mathbb{R}^N$ at a given time $t$, the reservoir dynamics at the next time step evolves as: \begin{equation}\label{eq1}
    \boldsymbol{r}_{t+1} = f(\boldsymbol{W}_{in} \boldsymbol{u}_t + \boldsymbol{W}_r \boldsymbol{r}_t + \boldsymbol{b}),
\end{equation}
where $\boldsymbol{W}_{in}$ is the input matrix mapping input data to the neuron domain, $\boldsymbol{W}_r$ is the interconnection matrix between neurons, $\boldsymbol{b}$ is the bias vector, and $f$ is the activation function that is typically nonlinear. To further control the memory of RC, many architectures also incorporate an additional hyperparameter, know as the leaking rate, to balance the current nonlinear activation with the previous reservoir state. After evolving the reservoir for a sufficient time based on Eq. \eqref{eq1}, a linear estimator can be trained to map the obtained reservoir states to the target outputs $\boldsymbol{o}_{t}$ by defining $
    \boldsymbol{\hat{o}}_{t} = \boldsymbol{W}_{out} \boldsymbol{r}_t \approx \boldsymbol{o}_{t}$, where $\boldsymbol{W}_{out}$ is a readout layer mostly optimized through analytic linear regression (see Methods). Note that in the time series prediction tasks, the desired output is often the input, i.e., $\boldsymbol{o}_{t}=\boldsymbol{u}_{t}$. After training, the reservoir can autonomously evolve along a trajectory by closing the feedback loop in forecasting tasks. Importantly, the fixed nature of $\boldsymbol{W}_{in}$ and $\boldsymbol{W}_r$ renders RC a hardware-agnostic computing framework. RC also bypasses the challenges encountered in previous RNN training algorithms, such as backpropagation through time \cite{werbos1990backpropagation}, as it only trains the readout matrix $\boldsymbol{W}_{out}$.

In contrast, the recently proposed NGRC builds the reservoir features directly from the input data in the polynomial form (Fig. \ref{Fig1}b). While the polynomial order and the number of delayed inputs in NGRC are flexible and task-dependent, we formulate the NGRC with up to quadratic terms and inputs from two time steps for simplicity: 
\begin{flalign}
\label{eq2}
\begin{split}
    & \boldsymbol{r}_{t+1} = (1, \underbrace{\overbrace{u_{1,t},...,u_{M,t}}^{\boldsymbol{u}_t^T},\overbrace{u_{1,t-k},...,u_{M,t-k}}^{\boldsymbol{u}_{t-k}^T}}_{\mathrm{Linear \thinspace terms}}, \\
    & \underbrace{\overbrace{u_{1,t}^2,...,u_{M,t}^2}^{\mathbb{U}(\boldsymbol{u}_t \otimes \boldsymbol{u}_{t})},\overbrace{u_{1,t-k}^2,...,u_{M,t-k}^2}^{\mathbb{U}(\boldsymbol{u}_{t-k} \otimes \boldsymbol{u}_{t-k})},\overbrace{u_{1,t}u_{1,t-k},...,u_{M,t}u_{M,t-k}}^{\mathbb{U}(\boldsymbol{u}_t \otimes \boldsymbol{u}_{t-k})}}_{\mathrm{Nonlinear \thinspace quadratic \thinspace terms}}),
\end{split}
\end{flalign}
where $1$ denotes the bias term, $\boldsymbol{u}_{t-k}\in \mathbb{R}^M$ is a delayed input from $k$ previous time steps ($k=1$ is used hereafter unless otherwise specified). $\otimes$ denotes the outer product and $\mathbb{U}$ is defined as an operation to collect all unique monomials from the matrix vectorization of the outer product of two vectors.

\begin{figure*}[!htp]
  \centering{
  \includegraphics[width = 0.88\linewidth]{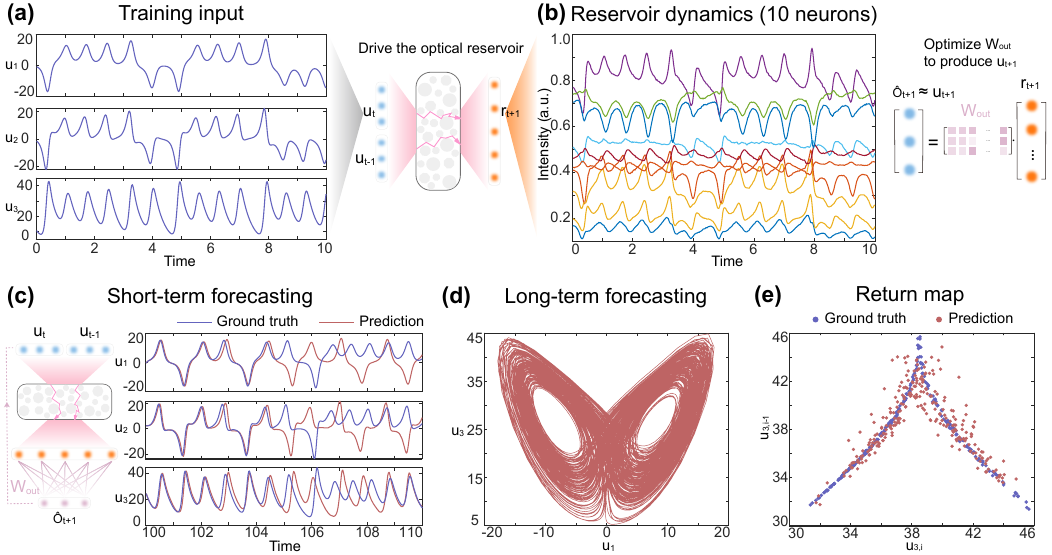}
  } 
    \caption{\noindent\textbf{Optical NGRC for Lorenz63 attractor forecasting.} \textbf{(a)} Time series of the Lorenz63 attractor (state variables $u_1, u_2, u_3$) that drives the optical NGRC. At each time step of the training phase, the input states from the current ($\boldsymbol{u}_t$) and the previous ($\boldsymbol{u}_{t-1}$) time steps are encoded to the optical system to generate reservoir features ($\boldsymbol{r}_{t+1}$).
    \textbf{(b)} The temporal evolution of 10 randomly selected optical reservoir nodes (out of 2,000 nodes), which resembles the dynamics of the input data. After training iterations of 4,000 time steps, a linear estimator $\boldsymbol{W}_{out}$ is trained to match the weighted sums of the  reservoir features ($\boldsymbol{\hat{o}}_{t} = \boldsymbol{W}_{out}\boldsymbol{r}_{t}$) with the input data at the next time step ($\boldsymbol{u}_{t}$), i.e., $\boldsymbol{\hat{o}}_{t}\approx \boldsymbol{u}_{t}$.  \textbf{(c)} Once $\boldsymbol{W}_{out}$ is optimized, the optical NGRC is switched to the autonomous mode and experimentally predicts short-term results for 400 time steps. The normalized root mean square error (NRMSE) over the first 5 time units of the prediction phase is 0.0971. \textbf{(d)} The optical NGRC projects onto an attractor similar to the Lorenz63 attactor, experimentally obtained by the long-term forecasting results of 8,000 time steps. \textbf{(e)} The return map of the ground truth (blue) and the experimental prediction (red).
  }
  \label{Fig2}
\end{figure*} 

With these in mind, we now adapt NGRC into a format that is compatible with optical implementations (Figs. \ref{Fig1}(c)-(e)), such that we can use a similar optical setup as for conventional optical RC \cite{dong2019optical,rafayelyan2020large} for NGRC.
Our computing engine employs a continuous-wave laser as the light source, a phase-only SLM for data encoding, a scattering medium for information mixing, and a camera for feature detection (see Methods and Supplementary Note 1). 
Here, the input data from different time steps is encoded onto the spatial phase profile of light via the SLM. The scattering medium linearly connects the input and output optical fields via a transmission matrix, mixing the input as speckle patterns at the camera plane. Then, the formation of speckle feature vectors is analogous to random projection, which is a ubiquitous computation tool widely used in mathematics and signal processing \cite{gigan2022imaging}. Taking into account the nonlinear responses of phase encoding of the SLM ($x \rightarrow{\mathrm{exp}(ix)}$) and square-law detection of the camera ($x \rightarrow |x|^2$), the overall optical process defines the nonlinear mapping between the inputs ($\boldsymbol{u}_t$ and $\boldsymbol{u}_{t-1}$) and the reservoir state ($\boldsymbol{r}_{t+1}$) as:
\begin{equation}\label{eq3}
    \boldsymbol{r}_{t+1} = |\boldsymbol{W}_{in1} \mathrm{exp}(i\boldsymbol{u}_t) + \boldsymbol{W}_{in2} \mathrm{exp} (i\boldsymbol{u}_{t-1}) + \boldsymbol{b}|^2,
\end{equation}
where $\boldsymbol{W}_{in1}$ and $\boldsymbol{W}_{in2}$ are random complex matrices given by the optical scattering medium. In contrast to conventional optical RC schemes where the reservoir state at the time step $t+1$ is calculated based on the current input $\boldsymbol{u}_t$ and the reservoir state $\boldsymbol{r}_t$ \cite{paquot2012optoelectronic,larger2012photonic,martinenghi2012photonic,brunner2013parallel,vinckier2015high,duport2016fully,larger2017high,antonik2019human,dong2019optical,rafayelyan2020large}, we replace $\boldsymbol{r}_t$ with the delayed input $\boldsymbol{u}_{t-1}$ (Fig. \ref{Fig1}(d)).
Notably, this is different from the conventional RC framework augmented by delayed inputs \cite{Jauriguee23121560,Jaurigue_2024}. Such a modification generates implicitly the polynomial forms of input variables at time steps $t$ and $t-1$ (Fig. \ref{Fig1}(e)), as evident by expanding $\boldsymbol{r}_{t+1}$ via Taylor series decomposition (see Supplementary Note 1):
\begin{flalign}
\label{eq4}
\begin{split}
    &\boldsymbol{r}_{t+1} \approx \boldsymbol{M}_s \cdot [1, \underbrace{\boldsymbol{u}_t^T,\boldsymbol{u}_{t-1}^T}_{\mathrm{Linear \thinspace terms}}, \\
    &\underbrace{\mathbb{U}(\boldsymbol{u}_{t} \otimes \boldsymbol{u}_{t}), \mathbb{U}(\boldsymbol{u}_{t-1} \otimes \boldsymbol{u}_{t-1}), \mathbb{U}(\boldsymbol{u}_{t} \otimes \boldsymbol{u}_{t-1})}_{\mathrm{Quadratic \thinspace terms}}, ...]^T,
\end{split}
\end{flalign}
where $\boldsymbol{M}_s$ is a matrix given by the optical system, which mixes the underlying polynomial terms ($\boldsymbol{\Theta_t}$) embedded in the speckle feature vector. In essence, the speckle vector can be understood as weighted sums of linear, quadratic and higher-order polynomial terms of $\boldsymbol{u}_t$ and $\boldsymbol{u}_{t-1}$. Stated differently, our optical system can compute similar feature terms just as the NGRC does in Eq. \eqref{eq2}, only that an additional matrix linearly couples all these explicit terms together. Besides, the optimized linear readout matrix $\boldsymbol{W}_{out}$ trained in optical NGRC can be related to $\boldsymbol{W}_{out}^\prime$ in digital NGRC by $\boldsymbol{W}_{out}^\prime \approx\boldsymbol{W}_{out} \boldsymbol{M}_s$, thus validating that our optical implementation is equivalent to the digital NGRC operation (see Supplementary Note 1). Due to the presence of the system-given matrix $\boldsymbol{M}_s$, the optical implementation operates in an indirect manner. This indirect way of implementing NGRC optically also inherits the advantage of interpretability of digital NGRC, as we can explain the reservoir computations via the synthesized features of the time-delayed inputs and then a readout layer $\boldsymbol{W}_{out}^\prime$ linearly combining them for specific tasks.

\vspace{0.1cm}

\noindent\textbf{Forecasting Lorenz attractor.} To demonstrate the effectiveness of the proposed optical NGRC, we firstly apply our setup to the low-dimensional Lorenz63 time series forecasting task (see Methods for dataset information). As illustrated in Fig. \ref{Fig2}, we initially drive the optical system by encoding $[\boldsymbol{u}_t, \boldsymbol{u}_{t-1}]^T$ with a time interval of $\Delta t = 0.025$ onto the SLM and we gather in total 4,000 reservoir speckle feature vectors used for training (see Methods for experimental details). Figure \ref{Fig2}(b) showcases the dynamics of 10 reservoir neurons measured in the experiment, providing nonlinear representations that reflect the characteristics of the input dataset. The smoothness of the reservoir dynamics, essential for reliable RC training \cite{hu2023tackling}, is guaranteed by the high stability of our experimental setup (see Supplementary Note 2). Then, we regress a digital readout layer $\boldsymbol{W}_{out}$ to map the reservoir state $\boldsymbol{r}_{t}$ to the next time step in the Lorenz63 attractor, i.e., $\boldsymbol{\hat{o}}_{t} = \boldsymbol{W}_{out} \boldsymbol{r}_{t} \approx \boldsymbol{u}_{t}$ (see Methods for training details). After $\boldsymbol{W}_{out}$ is obtained, the optical NGRC is used as an autonomous dynamical system for predicting another 400 time steps (see Supplementary Algorithm 1). The training and prediction are implemented in the optical experiment with an effective system frame rate of around 10 Hz (see Supplementary Note 2). 

In the short term, the optical NGRC shows decent forecasting capability of the Lorenz63 time series up to around 4 time units (Fig. \ref{Fig2}(c)). Note that due to the nature of the chaotic systems, the prediction by optical NGRC would eventually diverge after a certain period of time, just as all models predicting chaos. Such a divergence does not imply the collapse of the RC model, rather, the ergodic (statistical) properties of the attractor are still preserved by RC, known as `climate' replication \cite{pathak2017using,lu2018attractor}. To this end, we run the trained optical NGRC for an extended period of 8,000 time steps. The long-term prediction consistently reproduces the manifold, as evident by the phase-space trajectory with double wings shown in Fig. \ref{Fig2}(d). Beyond visual inspection, we quantitatively evaluate the long-term forecasting performance by calculating the return map, in which the successive maxima of the third dimension $\boldsymbol{u}_3$ in time are collected and plotted. As shown in Fig. \ref{Fig2}(e), the experimentally obtained data points collectively cluster around the ground truth curve, albeit with a deviation due to the presence of the experimental noise and quantization.

\vspace{0.1cm}

\begin{figure*}[!htp]
  \centering{
  \includegraphics[width = 0.88\linewidth]{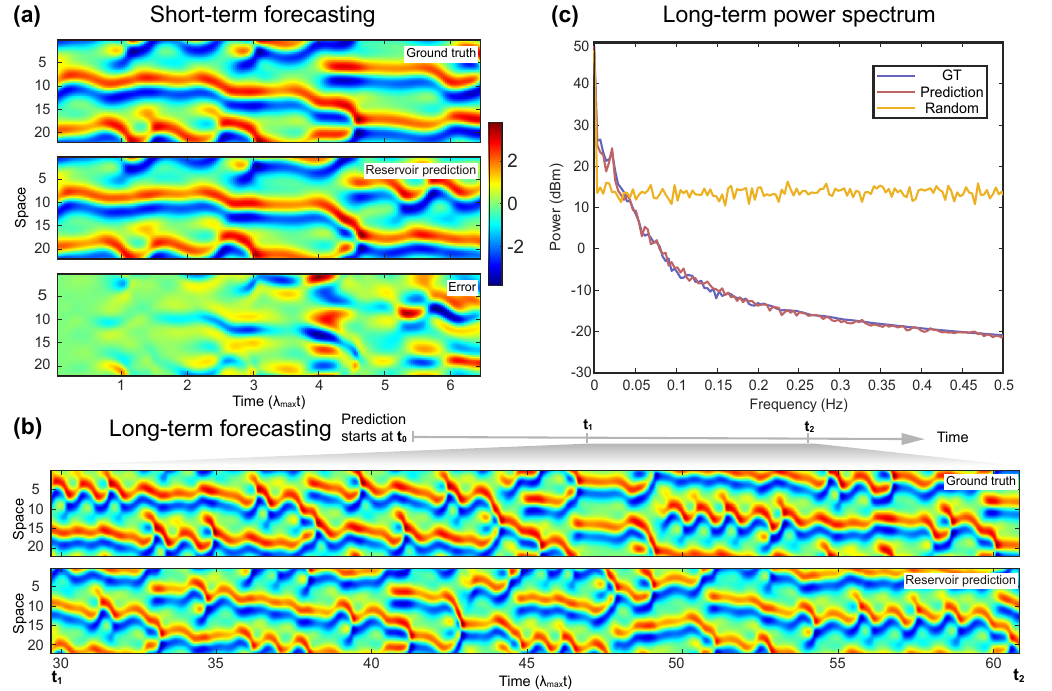}
  } 
    \caption{\noindent\textbf{Optical NGRC for Kuramoto-Sivashinsky time series forecasting.} \textbf{(a)} Experimental short-term prediction results of the Kuramoto-Sivashinsky (KS) time series with a domain size of $L=22$ and a spatial sampling of $S=64$. An optical NGRC with 2,500 optical reservoir nodes is used for KS forecasting, which employs the current ($\boldsymbol{u}_t$) and the previous ($\boldsymbol{u}_{t-1}$) time steps in each training iteration for a total training length of  6,000 time steps.
    The error subfigure (bottom) is the element-wise difference between the ground truth (top) and the experimental prediction (middle). 
    The temporal axis is normalized by its largest Lyapunov time ($\lambda_{max}=0.043$).
    \textbf{(b)} A part of the long-term prediction results by optical NGRC (between $t_1$ and $t_2$, where the prediction starts at $t_0$). Albeit the  complete deviation between the KS ground truth (top) and the optical NGRC predicted output (bottom) at the element-wise level, the optical NGRC replicates the long-term behavior of the KS chaotic system. 
    \textbf{(c)} The power spectra of the long-term prediction in \textbf{(b)} (red), the KS ground truth (blue) and a random noise signal (yellow). The power spectra of the ground truth and optical NGRC predictions are in good agreement, in stark contrast to the power spectrum of the random noise background.
  }
  \label{Fig3}
\end{figure*} 
\noindent\textbf{Forecasting Kuramoto-Sivashinsky time series.} Next, we use the optical NGRC in a more challenging scenario by forecasting another standard benchmark dataset in RC, i.e., a large-scale spatiotemporal chaotic KS time series (see Methods for dataset information).  In Fig. \ref{Fig3}(a), we illustrate the short-term prediction results obtained in the experiment through online Bayesian optimization (see Methods and Supplementary Note 2). After training the optical NGRC, the optical system can forecast the KS system reasonably well up to around 4 Lyapunov times (see definition in Methods), longer than the 2.5 Lyapunov times achieved previously \cite{rafayelyan2020large}. At the prediction phase, the normalized root mean square (NRMSE) over the test period (6.45 Lyapunov times) is calculated as 0.2988 (see definition in Methods). We remark that, conventional RC typically necessitates a quite long warm-up period ranging from $100$ to $100,000$ time steps \cite{lu2018attractor,griffith2019forecasting}, which can be challenging in situations where training data is limited or the physical evolution of the reservoir system is time-consuming. Thanks to the NGRC operation \cite{gauthier2021next}, we use in this study only 2 time steps for warm up and 6,000 time steps for training, much shorter than the total 90,500 time steps used in ref. \cite{rafayelyan2020large}. Taken together, the better prediction performance in optical NGRC, combined with much less warm up and training data as well as a smaller reservoir size, collectively suggest the superiority of the optical NGRC over the conventional optical RC based on scattering media. The performance improvement using optical NGRC is also confirmed by numerical simulations (see Supplementary Note 3). 

\begin{figure*}[!htp]
  \centering{
  \includegraphics[width = 0.88\linewidth]{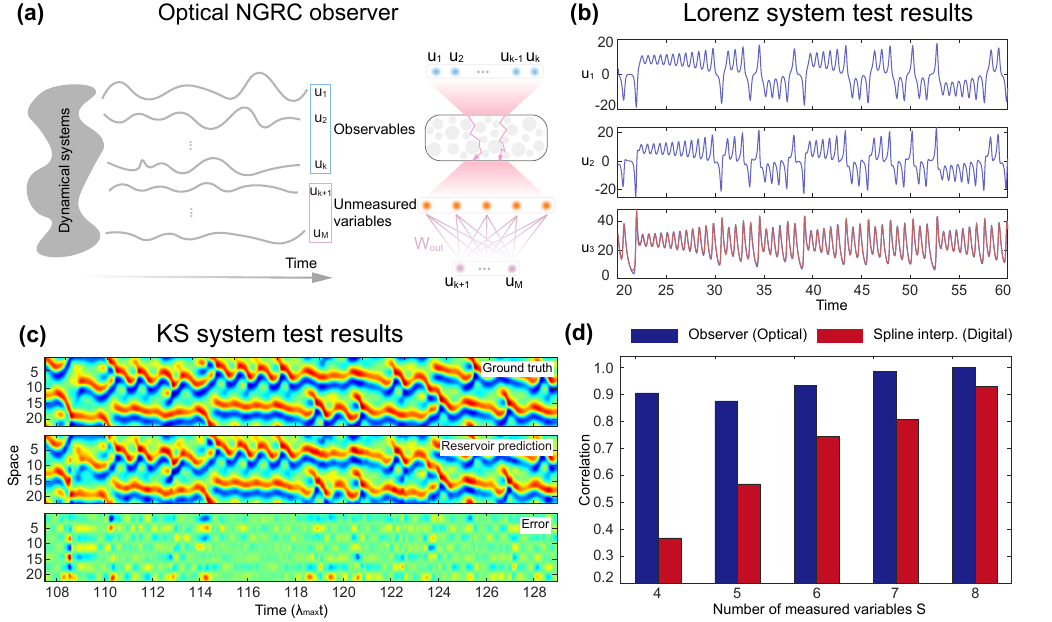}
  } 
    \caption{\noindent\textbf{Optical NGRC observer.} \textbf{(a)} For a dynamical system, often partial information of the full state of the system is measurable, e.g., state variables $[u_1,...,u_k]^T$ are observables while $[u_{k+1},...,u_M]^T$ are unmeasured. The optical NGRC extracts information from measured observables (blue) and predicts unmeasured variables (purple) based on the state of the reservoir (orange).
    \textbf{(b)} Two variables $u_1$ and $u_2$ (blue) of the Lorenz63 system are provided as observables to infer the third variable $u_3$. The predicted output by optical NGRC observer (red) matches the ground truth (blue) with high accuracy (NRMSE $=$ 0.0169).
    \textbf{(c)} The optical NGRC observer results of the KS time series. 7 out of 64 spatial grids (evenly spaced in the spatial dimension) are input of the optical NGRC to infer the remaining 57 unmeasured variables. Top: ground truth; Middle: reservoir prediction (also including the observables for clarity); Bottom: error. \textbf{(d)} Performance comparison of the optical NGRC observer and the spline interpolation on the KS time series. The Pearson correlation between the optical NGRC observer prediction and the ground truth is consistently higher than that between the spline interpolation and the ground truth.
  }
  \label{Fig4}
\end{figure*} 

 Regarding its long-term prediction performance, we illustrate in Fig. \ref{Fig3}(b) a section (spanning from $t_1$ to $t_2$) of predicted outputs for 10,000 time steps (starting at $t_0$)  beyond the short-term regime. While the prediction completely deviates from the ground truth at the element-wise level, the visual inspection indicates that the optical NGRC captures the correct `climate' \cite{pathak2017using}. We substantiate this observation by quantitatively analyzing the power spectra of the predicted outputs, the KS ground truth and a random noise signal in Fig. \ref{Fig3}(c) (see data processing details in Methods). The long-term prediction results presented in Figs. \ref{Fig2} and \ref{Fig3} indicate that the optical NGRC effectively synchronizes with host prototypical systems, functioning as a physical twin without knowing their models.

\vspace{0.1cm}

\noindent\textbf{Optical NGRC observer.} We now proceed to the application of optical NGRC in a third benchmark task, referred to as the `reservoir observer' \cite{lu2017reservoir,gauthier2021next}. As illustrated in Fig. \ref{Fig4}(a), in many contexts when studying a dynamical system, it is common to have access to only a partial set of its complete degrees of freedom at a given time. An `observer' aims to deduce unmeasured variables from the measured ones (i.e., observables), for example, $[u_1,...,u_k]^T \xrightarrow{\mathrm{optical \thinspace NGRC}} [u_{k+1},...,u_M]^T$  (see Supplementary Algorithm 2). As before, we first train the optical reservoir in a supervised fashion, based on the limited number of time measurements where the full state variables of the system $[u_1,...,u_k,...,u_M]^T$ are accessible. Here, we conduct optical NGRC observer experiments on both the Lorenz63 and KS systems. 

To follow the convention of the observer task in digital NGRC \cite{gauthier2021next}, we employ one current and three delayed inputs to infer unmeasured variables, rather than the two inputs used in previous autonomous forecasting tasks. These four inputs are uniformly sampled with a stride of five time steps (see details in Supplementary Algorithm 2). For the Lorenz63 system, we infer $u_3$ from $u_1$ and $u_2$. Figure \ref{Fig4}(b) shows that only a short period of training time with 400 time steps yields decent predictions for $20<t<60$, manifesting the feasibility of our optical NGRC in this application. Going beyond, we investigate the optical NGRC observer in the spatiotemporal KS system, based on the sparse spatial information that is available. Specifically, for the domain size of $L=22$ of the KS time series studied in this work, we sample $64$ spatial points at each time step. The experiments are performed using the knowledge of uniformly sampled $S$ spatial state variables to infer the remaining set of $64-S$ variables. Figure \ref{Fig4}(c) presents the test results when $S=7$. We also vary $S$ from 4 to 8 and summarize the calculated correlation between the experimental results and the corresponding ground truth in Fig. \ref{Fig4}(d). To provide a digital baseline, the cubic spline interpolation method is also implemented (see data processing and performance metrics details in Methods). We observe that the optical NGRC observer consistently outperforms the spline interpolation, serving as an effective means to reconstruct unmeasured dynamical system variables.

%%%%%%%%%%%%%%%%%%%%%%%%%%%%%%%%%%%%%%%%%%%
\section*{Discussion}
\noindent{Generally}, for physical RC systems, experimental factors such as device quantization and noise can limit the RC performance. Nevertheless, these requirements are not overly stringent for optical NGRC. In Supplementary Note 4, we show that 7- or 8-bit depth SLM and camera are sufficient for the tasks demonstrated in this work. To reduce the effect of the noise, we optimize the setup and average the experimental measurements when deriving the reservoir states (see Supplementary Note 2). Beyond forecasting chaotic time series, it is interesting to explore other machine learning tasks with optical NGRC, such as speech recognition \cite{pedrelli2021hierarchical} and graph classification \cite{gallicchio2017deep,gallicchio2020fast}. At the architecture level, the optical NGRC can also be chained to create more advanced neural networks like deep or parallel NGRC \cite{gallicchio2017deep,barbosa2022learning} for enhanced expressivity and versatile functionalities \cite{kim2023neural}. For instance, we can employ a thin scattering medium to only introduce the random mixing locally \cite{delloye2023optical}, so as to implement parallel optical NGRC \cite{barbosa2022learning}.

It is important to note that many steps in our current implementation involve processing in a digital computer like many other optical RC systems. Upgrading the system to an all-optical version remains a promising challenge, which could further improve the energy efficiency and computing throughput. Currently, we employ only a small fraction of pixels of the encoding SLM and detection camera. Our optical NGRC system can be scaled to accommodate larger inputs and larger reservoir output dimensions, allowing us to explore the potential scaling advantages of optical computing with scattering media as discussed in refs. \cite{ohana2020kernel,rafayelyan2020large,pierangeli2021scalable}.
While edge computing hardware have shown excellent performance in implementing NGRC \cite{kent2024controlling}, it is restricted to accelerating relatively small-scale tasks and optical systems comprising more advanced components may offer further improvements (see Supplementary Note 5).

Compared to digital NGRC, the scaling of optical NGRC and consequently the computational complexity, exhibits notable distinctions. Specifically, digital NGRC features polynomial scaling with respect to the input data dimension (see Supplementary Note 5), provided that dimension reduction of input data is not performed based on any prior knowledge \cite{pyle2021domain,gauthier2021next}. This could impose a progressively heavier burden for digital NGRC to process large-scale systems. For example, for the same KS time series prediction studied in this work, the original digital NGRC scheme would require a reservoir size of 8,385 when considering up to second-order polynomial terms, while we only use 2,500 in optical NGRC (see Supplementary Note 5). More interestingly, due to the use of phase encoding and intensity detection in the optical NGRC, higher-order polynomials functions of delayed inputs (beyond the second-order and in principle infinite orders due to Taylor expansion) are naturally embedded in every speckle grain mode, and they can be readily accessed by the linear readout layer if needed. As such, we do not have to manually determine the needed polynomial order or select the terms as necessitated in digital NGRC. In our case, we may use the phase encoding range as another hyperparameter to optimize the reservoir.

In summary, we propose and experimentally demonstrate in this work an efficient optical NGRC scheme based on light scattering through disordered media. Similar to the spirit of digital NGRC, optical NGRC generates the mixture of polynomial functions of time-delayed inputs. We leverage these features embedded in optical speckles for various benchmark tasks in RC, ranging from short-term and long-term forecasting to reservoir observer in Lorenz63 and KS chaotic systems.
Optical NGRC features several advantages over conventional optical RC based on scattering media in significantly shorter training length, fewer hyperparameters, increased interpretability and greater expressivity, and may hold the prospect in scalability compared to digital NGRC towards large-scale chaotic systems (see Supplementary Table S1). Broadly, the proposed optical NGRC framework is hardware-agnostic, which could inspire new possibilities for a wide variety of physical RC substrates.

\vspace{0.1cm}

During the finalization of the manuscript, we became aware that a related work on optical NGRC observer was uploaded to arXiv \cite{cox2024photonic}.

\vspace{0.5cm}

\noindent\textbf{Methods}
\medskip
\begin{footnotesize}

\noindent\textbf{Experimental setup.} The optical NGRC system (Supplementary Fig. 1) is primarily composed of a continuous-wave laser, an SLM, a disordered medium and a camera. The light from a low-power (2.5 mW) polarization-maintaining laser at 635 nm (Thorlabs, S1FC635PM) is delivered to a pinhole via a fiber.
After the free-space propagation for a diffraction length of 100 mm from the pinhole, the input laser beam is collimated by a lens (L1, $f = 100$ mm). A polarizing beam splitter is used to match the output beam polarization with the working-axis of the following reflective phase-only SLM (Meadowlarks, HSP512L-1064). The input states are encoded onto the spatial wavefront of the laser beam. The modulated beam then passes through a $4-f$  relay system (L2, $f = 100$ mm; L3, $f = 100$ mm) to reach the front surface of the scattering medium. In the experiment, we use a ground glass diffuser as the scattering medium, which is prepared by sandblasting the surface ($\phi$ 22 mm) of a microscope coverslip (1.5H, $\phi$ 25 mm, Deckgläser) with 220 grit white fused alumina. The full width at half maximum (FWHM) scattering angle of the diffuser is approximately 10 degrees. After the scattering process,  the laser beam propagates freely for a length of 125 mm. The combined effects of multiple scattering and free-space propagation generate the reservoir states containing rich information of the inputs. The reservoir states are in the form of speckle patterns and are captured by a CMOS camera (Basler, acA1920-40um).

\vspace{0.1cm}

\noindent\textbf{Lorenz63 attractor, Kuramoto-Sivashinsky equation and Lyapunov exponents.} 
The Lorenz63 attractor is a canonical chaotic manifold representing a simplified model of a weather system proposed by Lorenz in 1963 \cite{lorenz1963deterministic}, described by three ordinary differential equations:
\begin{equation}\label{eq5}
\begin{aligned}
    &\dot{u_1} = \sigma(u_2-u_1), \\
    &\dot{u_2} = u_1 (\rho - u_3) - u_2, \\
    &\dot{u_3} = u_1 u_2 - \beta u_3,
\end{aligned}
\end{equation}
where $\sigma$, $\rho$, and $\beta$ determine the system dynamics and $[u_{1,t},u_{2,t},u_{3,t}]^T$ is the system state variables at time $t$. In this work, we use the parameters $\sigma=10$, $\rho=28$ and $\beta=8/3$, which gives rise to rich and chaotic dynamics that evolves on the double-wing attractor in the phase space. We integrate the equations using a fourth-fifth order Runge-Kutta method with a time step of $\Delta t = 0.05$.

 The Kuramoto-Sivashinsky equation is a partial differential equation that models many nonlinear systems with intrinsic instabilities, such as hydrodynamic turbulence and wave propagation in chemical reaction-diffusion systems \cite{hohenberg1989chaotic}. In this equation, dynamics at different scales interact mutually to generate spatiotemporal complexity governed by:
\begin{equation}\label{eq6}
\partial_t u + \partial_x^4 u + \partial_x^2 u + u \partial_x u=0,
\end{equation}
where the field $u(x,t)$ is periodic on the spatial domain $0\leq x < L$, that is $u(x,t)=u(x+L,t)$ with $L$ representing the spatial domain size. As the domain size $L$ increases, the KS evolution changes rapidly. We use $L=22$ in this study, which offers sufficient chaotic dynamics. We integrate the system based on a fourth order time-stepping method, on a spatial sampling grid of 64 ($S=64$) and a time step of $\Delta t=0.25$.

The knowledge of Lyapunov exponents represents the most basic yet pervasive measure of a dynamical system. In simple terms, a (global) Lyapunov exponent is the average rate at which the system diverges from its initial point in the phase space along one degree of freedom. Therefore, high-dimensional systems contain multiple Lyapunov exponents, collectively forming a Lyapunov spectrum. To calculate the spectrum, we initialize multiple orthogonal vectors in different directions as perturbations and evaluate their average divergences along evolution compared to the dynamics without perturbations \cite{edson2019lyapunov}.  In particular, the largest Lyapunov exponent $\lambda_{max}$ serves as an effective indicator to evaluate whether the system exhibits chaotic behavior ($\lambda_{max} > 0$) or non-chaotic behavior ($\lambda_{max} < 0$). Multiplying time by $\lambda_{max}$ yields the Lyapunov time in Figs. \ref{Fig3}-\ref{Fig4}, which denotes the average duration for errors to grow by a factor of $e$. For the Lorenz63 attractor, $\lambda_{max}=0.91$. For the KS equation studied in this work, $\lambda_{max}=0.043$.

\vspace{0.1cm}
\noindent\textbf{Data processing in the experiments.} Here we provide more details of data encoding and processing used in this study. First, we normalize the Lorenz63 and KS time series to the range of $[0,1]$ with respect to their global minimum and maximum values. In the experiments, we linearly scale the normalized input data to $[0,\pi]$ and encode it to the phase of light via an SLM. 
For short-term prediction, we apply two consecutive time steps, i.e., the current and previous inputs $\boldsymbol{u}_t$ and $\boldsymbol{u}_{t-1}$, to forecast the next step evolution ($\boldsymbol{u}_{t+1}$), and we introduce a relative weight $\eta$ between these two inputs as a hyperparameter to potentially improve the performance. With another bias hyperparameter $b$, the input vector to be sent to the SLM is written as $\pi [\boldsymbol{u}_t,\eta \boldsymbol{u}_{t-1},b]^T$. 
We remark that in this work we mostly use the phase range of $[0, \pi$] of the SLM, which results in an effective bit depth of 7 bits (0 to 127 in grayscale) for the encoding SLM. Practically, to reduce the crosstalk between pixels on the SLM, we use multiple pixels (macropixel) to represent each element of the input data vector above. Different macropixel sizes are used depending on the data dimensions (see Supplementary Table S2). In cases when the central region of the SLM is not entirely utilized, the unmodulated pixels serve as a static bias. To remove the unmodulated background light and unused periphery pixels from the SLM, we superimpose a blazed grating mask over the encoded data mask, and select the first-order diffraction at the Fourier plane of the $4-f$ system. We capture the speckle patterns within a predefined region of interest by the camera and downsample the measured images at intervals matching the speckle grain size, which is determined through speckle auto-correlation analysis. Subsequently, the speckle images are normalized from a range of 0 to 255 (8 bits) to a range of 0 to 1. We then randomly select independent nodes as needed from the normalized image, and flatten them into a reservoir feature vector for the following digital readout layer. To improve the forecasting performance, we concatenate the reservoir state and the current input state for prediction as applied in previous works \cite{dong2020reservoir,rafayelyan2020large}. Afterwards, the predicted output is scaled back to the original data range based on the previously determined minimum and maximum values. We summarize all the aforementioned parameters used in the experiments in Supplementary Table S2.

Once sufficient training reservoir states are collected, we train a digital linear readout layer $\boldsymbol{W}_{out}$ by the Tikhonov regularization method to map the reservoir states $\boldsymbol{R}$ to the targets $\boldsymbol{O}$. In particular, the optimal $\boldsymbol{W}_{out}$ is computed through minimizing the following objective function:
\begin{equation}
    \boldsymbol{W}_{out}= {\rm argmin}(\Vert   \boldsymbol{W}_{out} \boldsymbol{R}- \boldsymbol{O}\Vert_2^2 + \beta\Vert \boldsymbol{W}_{out}\Vert_2^2),
\end{equation}
where $\beta$ is the ridge regularization parameter to punish large weight values in $\boldsymbol{W}_{out}$. The ridge regression can be computed efficiently via the explicit solution $\boldsymbol{W}_{out}=\boldsymbol{O}\boldsymbol{R}^T(\boldsymbol{R}\boldsymbol{R}^T+\beta \boldsymbol{I})^{-1} $, without the need of error backpropagation. $\beta$ is an important hyperparameter which can improve the generalization ability and avoid overfitting, especially when the number of reservoir nodes is larger than the training length. When searching for the optimal $\beta$ in the reservoir observer task, singular value decomposition of $\boldsymbol{R}$ can be used to further accelerate the computations.

For the quantitative analysis of the long-term power spectrum  reported in Fig. \ref{Fig3}c, we apply a sliding window approach similar to the short-time Fourier transform. Specifically, we sample one spatial grid point ($32^{\rm nd}$ is used) from the 64 spatial grids of the spatiotemporal data to create a one-dimensional time series. Then we partition the time series of 10,000 data points into 20 intervals, each comprising 500 data points. Subsequently, we calculate the corresponding power spectrum by Fourier transform for each interval, and average them over all intervals. In this way, we obtain smoother power spectra of the time series and avoid local oscillations. As for the random noise signal, we initialize it by drawing random numbers from a uniform distribution. In the figure, we only illustrate the positive frequency part of the power spectra since it is symmetric with the negative part.

To establish the digital baseline for the optical reservoir observer, we use the cubic spline interpolation method, which resorts to low-order polynomials for smooth and accurate fitting while mitigating high-order polynomial oscillations. To do so, we employ the \textit{CubicSpline} function from the SciPy Python library with a periodic boundary condition.

For short-term forecasting of chaotic time series, we use the online Bayesian optimization approach, i.e., we run the optical NGRC setup on-the-fly during the hyperparameter optimization. This is an effective approach to achieve stable and reliable predictions from analog reservoirs with noise \cite{yperman2016bayesian}. Compared to other hyperparameter optimization techniques such as grid search or random search, Bayesian optimization is recently founded to be the optimal approach due to its fast convergence and effectiveness, especially in scenarios with large search spaces \cite{yperman2016bayesian,griffith2019forecasting,shahriari2015taking}. Essentially, a probabilistic surrogate model is optimized to predict the optimal parameters based on observed data under given metrics. And it introduces randomness to explore new parameter spaces over iterations, which effectively decreases the risk of getting trapped in local minima. In practice, we typically run 20 to 30 iterations using the \textit{bayesOpt} library in MATLAB during experiments (see Supplementary Note 2).

For all experimental data collection, we use MATLAB software on a desktop equipped with an Intel(R) Core(TM) i7-6700 CPU and 32 GB RAM. For the data analysis and simulations, we use another desktop with an AMD EPYC 7351P CPU and 64GB RAM. 

\vspace{0.1cm}

\noindent\textbf{Performance evaluation metrics.} 
Here we describe two metrics used in the data analysis and performance evaluation. The metric NRMSE used in this work is defined as ${\rm NRMSE}=\frac{1}{O_{\rm max}} \sqrt{\frac{\sum_{i=1}^K \sum_{j=1}^P (\hat{o}_{j,i} - o_{j,i})^2}{KP} }$, where $O_{\rm max}$ represents the maximum value of the ground truth dataset $O$, $K$ is the total number of time steps and $P$ is the output size. This metric is useful to understand the overall performance across a certain period of the time series. When comparing the performance of optical reservoir observer with spline interpolation on the KS system, we use the Pearson correlation coefficient calculated as $r=\frac{\sum_{i=1}^K \sum_{j=1}^P (\hat{o}_{j,i} - \hat{O}_{\rm mean})(o_{j,i} - O_{\rm mean})}{\sqrt{[\sum_{i=1}^K \sum_{j=1}^P (\hat{o}_{j,i} - \hat{O}_{\rm mean})^2] [\sum_{i=1}^K \sum_{j=1}^P (o_{j,i} - O_{\rm mean})^2]}}$, with $\hat{O}_{\rm mean} (O_{\rm mean})$  denoting the mean value of the predicted outputs (ground truth).

\vspace{0.1cm}

\vspace{0.1cm}

\noindent \textbf{Data and code availability}: 
The data and codes that support
the plots and reservoir computing simulations within the paper are
available at 
\href{https://github.com/comediaLKB/Optical-NGRC-based-on-multiple-light-scattering}{https://github.com/comediaLKB/Optical-NGRC-based-on-multiple-light-scattering}.
\vspace{0.1cm}

\noindent \textbf{Acknowledgements}: The authors acknowledge helpful discussions with Jonathan Dong and acknowledge Daniel J. Gauthier for constructive comments on the manuscript. This work was supported by Swiss National Science Foundation (SNF) projects LION and APIC (TMCG-2$\_$213713), ERC SMARTIES and Institut Universitaire de France. H.W. acknowledges China Scholarship Council and National Natural Science Foundation of China (623B2064 and 62275137). J. H. acknowledges SNF fellowship (P2ELP2$\_$199825). Y.B. acknowledges the support from Basic Science Research Program through the National Research Foundation of Korea (NRF) funded by the Ministry of Education (2022R1A6A3A03072108) and European Union’s Horizon Europe research and innovation programme (N°101105899). Q.L. acknowledges National Natural Science Foundation of China (62275137). \vspace{0.1cm}

\noindent \textbf{Author contributions}: J.H. and H.W. conceived the project. H.W, J. H. and Y.B. developed the optical setup and performed the experiments. H.W., J.H., and K.T. performed the simulations. H.W. and J.H. analyzed the results and wrote the manuscript with inputs from all authors. S.G., Q.L. and J.H. supervised the project.
\vspace{0.1cm}

\noindent \textbf{Competing interests}: The authors declare that they have no competing interests.
\end{footnotesize}

\newpage
\renewcommand{\bibpreamble}{
$^\ast$These authors contributed equally to this work.\\
$^\dagger${Corresponding authors: \textcolor{magenta}{jianqi.hu@epfl.ch}, \textcolor{magenta}{qiangliu@tsinghua.edu.cn}, \textcolor{magenta}{sylvain.gigan@lkb.ens.fr}}\\
}

\bibliographystyle{naturemag}
\bibliography{ref}

\begin{thebibliography}{10}
\expandafter\ifx\csname url\endcsname\relax
  \def\url#1{\texttt{#1}}\fi
\expandafter\ifx\csname urlprefix\endcsname\relax\def\urlprefix{URL }\fi
\providecommand{\bibinfo}[2]{#2}
\providecommand{\eprint}[2][]{\url{#2}}

\bibitem{werbos1990backpropagation}
\bibinfo{author}{Werbos, P.~J.}
\newblock \bibinfo{title}{Backpropagation through time: what it does and how to do it}.
\newblock \emph{\bibinfo{journal}{Proceedings of the IEEE}} \textbf{\bibinfo{volume}{78}}, \bibinfo{pages}{1550--1560} (\bibinfo{year}{1990}).

\bibitem{mante2013context}
\bibinfo{author}{Mante, V.}, \bibinfo{author}{Sussillo, D.}, \bibinfo{author}{Shenoy, K.~V.} \& \bibinfo{author}{Newsome, W.~T.}
\newblock \bibinfo{title}{Context-dependent computation by recurrent dynamics in prefrontal cortex}.
\newblock \emph{\bibinfo{journal}{Nature}} \textbf{\bibinfo{volume}{503}}, \bibinfo{pages}{78--84} (\bibinfo{year}{2013}).

\bibitem{dambre2012information}
\bibinfo{author}{Dambre, J.}, \bibinfo{author}{Verstraeten, D.}, \bibinfo{author}{Schrauwen, B.} \& \bibinfo{author}{Massar, S.}
\newblock \bibinfo{title}{Information processing capacity of dynamical systems}.
\newblock \emph{\bibinfo{journal}{Scientific Reports}} \textbf{\bibinfo{volume}{2}}, \bibinfo{pages}{514} (\bibinfo{year}{2012}).

\bibitem{hu2023tackling}
\bibinfo{author}{Hu, F.} \emph{et~al.}
\newblock \bibinfo{title}{Tackling sampling noise in physical systems for machine learning applications: Fundamental limits and eigentasks}.
\newblock \emph{\bibinfo{journal}{Physical Review X}} \textbf{\bibinfo{volume}{13}}, \bibinfo{pages}{041020} (\bibinfo{year}{2023}).

\bibitem{6789852}
\bibinfo{author}{Maass, W.}, \bibinfo{author}{Natschläger, T.} \& \bibinfo{author}{Markram, H.}
\newblock \bibinfo{title}{Real-time computing without stable states: A new framework for neural computation based on perturbations}.
\newblock \emph{\bibinfo{journal}{Neural Computation}} \textbf{\bibinfo{volume}{14}}, \bibinfo{pages}{2531--2560} (\bibinfo{year}{2002}).

\bibitem{jaeger2001echo}
\bibinfo{author}{Jaeger, H.}
\newblock \bibinfo{title}{The “echo state” approach to analysing and training recurrent neural networks-with an erratum note}.
\newblock \emph{\bibinfo{journal}{Bonn, Germany: German National Research Center for Information Technology GMD Technical Report}} \textbf{\bibinfo{volume}{148}}, \bibinfo{pages}{13} (\bibinfo{year}{2001}).

\bibitem{jaeger2004harnessing}
\bibinfo{author}{Jaeger, H.} \& \bibinfo{author}{Haas, H.}
\newblock \bibinfo{title}{Harnessing nonlinearity: Predicting chaotic systems and saving energy in wireless communication}.
\newblock \emph{\bibinfo{journal}{Science}} \textbf{\bibinfo{volume}{304}}, \bibinfo{pages}{78--80} (\bibinfo{year}{2004}).

\bibitem{Sussillo_Abbott_2009}
\bibinfo{author}{Sussillo, D.} \& \bibinfo{author}{Abbott, L.}
\newblock \bibinfo{title}{Generating coherent patterns of activity from chaotic neural networks}.
\newblock \emph{\bibinfo{journal}{Neuron}} \textbf{\bibinfo{volume}{63}}, \bibinfo{pages}{544–557} (\bibinfo{year}{2009}).

\bibitem{yan2024emerging}
\bibinfo{author}{Yan, M.} \emph{et~al.}
\newblock \bibinfo{title}{Emerging opportunities and challenges for the future of reservoir computing}.
\newblock \emph{\bibinfo{journal}{Nature Communications}} \textbf{\bibinfo{volume}{15}}, \bibinfo{pages}{2056} (\bibinfo{year}{2024}).

\bibitem{pathak2018model}
\bibinfo{author}{Pathak, J.}, \bibinfo{author}{Hunt, B.}, \bibinfo{author}{Girvan, M.}, \bibinfo{author}{Lu, Z.} \& \bibinfo{author}{Ott, E.}
\newblock \bibinfo{title}{Model-free prediction of large spatiotemporally chaotic systems from data: A reservoir computing approach}.
\newblock \emph{\bibinfo{journal}{Physical Review Letters}} \textbf{\bibinfo{volume}{120}}, \bibinfo{pages}{024102} (\bibinfo{year}{2018}).

\bibitem{bollt2021explaining}
\bibinfo{author}{Bollt, E.}
\newblock \bibinfo{title}{On explaining the surprising success of reservoir computing forecaster of chaos? the universal machine learning dynamical system with contrast to var and dmd}.
\newblock \emph{\bibinfo{journal}{Chaos: An Interdisciplinary Journal of Nonlinear Science}} \textbf{\bibinfo{volume}{31}} (\bibinfo{year}{2021}).

\bibitem{pammi2023extreme}
\bibinfo{author}{Pammi, V.}, \bibinfo{author}{Clerc, M.}, \bibinfo{author}{Coulibaly, S.} \& \bibinfo{author}{Barbay, S.}
\newblock \bibinfo{title}{Extreme events prediction from nonlocal partial information in a spatiotemporally chaotic microcavity laser}.
\newblock \emph{\bibinfo{journal}{Physical Review Letters}} \textbf{\bibinfo{volume}{130}}, \bibinfo{pages}{223801} (\bibinfo{year}{2023}).

\bibitem{bianchi2020reservoir}
\bibinfo{author}{Bianchi, F.~M.}, \bibinfo{author}{Scardapane, S.}, \bibinfo{author}{L{\o}kse, S.} \& \bibinfo{author}{Jenssen, R.}
\newblock \bibinfo{title}{Reservoir computing approaches for representation and classification of multivariate time series}.
\newblock \emph{\bibinfo{journal}{IEEE Transactions on Neural Networks and Learning Systems}} \textbf{\bibinfo{volume}{32}}, \bibinfo{pages}{2169--2179} (\bibinfo{year}{2020}).

\bibitem{lu2017reservoir}
\bibinfo{author}{Lu, Z.} \emph{et~al.}
\newblock \bibinfo{title}{Reservoir observers: Model-free inference of unmeasured variables in chaotic systems}.
\newblock \emph{\bibinfo{journal}{Chaos: An Interdisciplinary Journal of Nonlinear Science}} \textbf{\bibinfo{volume}{27}} (\bibinfo{year}{2017}).

\bibitem{kim2021teaching}
\bibinfo{author}{Kim, J.~Z.}, \bibinfo{author}{Lu, Z.}, \bibinfo{author}{Nozari, E.}, \bibinfo{author}{Pappas, G.~J.} \& \bibinfo{author}{Bassett, D.~S.}
\newblock \bibinfo{title}{Teaching recurrent neural networks to infer global temporal structure from local examples}.
\newblock \emph{\bibinfo{journal}{Nature Machine Intelligence}} \textbf{\bibinfo{volume}{3}}, \bibinfo{pages}{316--323} (\bibinfo{year}{2021}).

\bibitem{antonelo2008event}
\bibinfo{author}{Antonelo, E.~A.}, \bibinfo{author}{Schrauwen, B.} \& \bibinfo{author}{Stroobandt, D.}
\newblock \bibinfo{title}{Event detection and localization for small mobile robots using reservoir computing}.
\newblock \emph{\bibinfo{journal}{Neural Networks}} \textbf{\bibinfo{volume}{21}}, \bibinfo{pages}{862--871} (\bibinfo{year}{2008}).

\bibitem{nakajima2021reservoir}
\bibinfo{author}{Nakajima, K.} \& \bibinfo{author}{Fischer, I.}
\newblock \emph{\bibinfo{title}{Reservoir Computing}} (\bibinfo{publisher}{Springer}, \bibinfo{year}{2021}).

\bibitem{tanaka2019recent}
\bibinfo{author}{Tanaka, G.} \emph{et~al.}
\newblock \bibinfo{title}{Recent advances in physical reservoir computing: A review}.
\newblock \emph{\bibinfo{journal}{Neural Networks}} \textbf{\bibinfo{volume}{115}}, \bibinfo{pages}{100--123} (\bibinfo{year}{2019}).

\bibitem{appeltant2011information}
\bibinfo{author}{Appeltant, L.} \emph{et~al.}
\newblock \bibinfo{title}{Information processing using a single dynamical node as complex system}.
\newblock \emph{\bibinfo{journal}{Nature Communications}} \textbf{\bibinfo{volume}{2}}, \bibinfo{pages}{468} (\bibinfo{year}{2011}).

\bibitem{du2017reservoir}
\bibinfo{author}{Du, C.} \emph{et~al.}
\newblock \bibinfo{title}{Reservoir computing using dynamic memristors for temporal information processing}.
\newblock \emph{\bibinfo{journal}{Nature Communications}} \textbf{\bibinfo{volume}{8}}, \bibinfo{pages}{2204} (\bibinfo{year}{2017}).

\bibitem{torrejon2017neuromorphic}
\bibinfo{author}{Torrejon, J.} \emph{et~al.}
\newblock \bibinfo{title}{Neuromorphic computing with nanoscale spintronic oscillators}.
\newblock \emph{\bibinfo{journal}{Nature}} \textbf{\bibinfo{volume}{547}}, \bibinfo{pages}{428--431} (\bibinfo{year}{2017}).

\bibitem{grollier2020neuromorphic}
\bibinfo{author}{Grollier, J.} \emph{et~al.}
\newblock \bibinfo{title}{Neuromorphic spintronics}.
\newblock \emph{\bibinfo{journal}{Nature Electronics}} \textbf{\bibinfo{volume}{3}}, \bibinfo{pages}{360--370} (\bibinfo{year}{2020}).

\bibitem{cai2023brain}
\bibinfo{author}{Cai, H.} \emph{et~al.}
\newblock \bibinfo{title}{Brain organoid reservoir computing for artificial intelligence}.
\newblock \emph{\bibinfo{journal}{Nature Electronics}} \bibinfo{pages}{1--8} (\bibinfo{year}{2023}).

\bibitem{markovic2020physics}
\bibinfo{author}{Markovi{\'c}, D.}, \bibinfo{author}{Mizrahi, A.}, \bibinfo{author}{Querlioz, D.} \& \bibinfo{author}{Grollier, J.}
\newblock \bibinfo{title}{Physics for neuromorphic computing}.
\newblock \emph{\bibinfo{journal}{Nature Reviews Physics}} \textbf{\bibinfo{volume}{2}}, \bibinfo{pages}{499--510} (\bibinfo{year}{2020}).

\bibitem{mehonic2022brain}
\bibinfo{author}{Mehonic, A.} \& \bibinfo{author}{Kenyon, A.~J.}
\newblock \bibinfo{title}{Brain-inspired computing needs a master plan}.
\newblock \emph{\bibinfo{journal}{Nature}} \textbf{\bibinfo{volume}{604}}, \bibinfo{pages}{255--260} (\bibinfo{year}{2022}).

\bibitem{jaeger2023toward}
\bibinfo{author}{Jaeger, H.}, \bibinfo{author}{Noheda, B.} \& \bibinfo{author}{Van Der~Wiel, W.~G.}
\newblock \bibinfo{title}{Toward a formal theory for computing machines made out of whatever physics offers}.
\newblock \emph{\bibinfo{journal}{Nature Communications}} \textbf{\bibinfo{volume}{14}}, \bibinfo{pages}{4911} (\bibinfo{year}{2023}).

\bibitem{stern2023learning}
\bibinfo{author}{Stern, M.} \& \bibinfo{author}{Murugan, A.}
\newblock \bibinfo{title}{Learning without neurons in physical systems}.
\newblock \emph{\bibinfo{journal}{Annual Review of Condensed Matter Physics}} \textbf{\bibinfo{volume}{14}}, \bibinfo{pages}{417--441} (\bibinfo{year}{2023}).

\bibitem{wetzstein2020inference}
\bibinfo{author}{Wetzstein, G.} \emph{et~al.}
\newblock \bibinfo{title}{Inference in artificial intelligence with deep optics and photonics}.
\newblock \emph{\bibinfo{journal}{Nature}} \textbf{\bibinfo{volume}{588}}, \bibinfo{pages}{39--47} (\bibinfo{year}{2020}).

\bibitem{shastri2021photonics}
\bibinfo{author}{Shastri, B.~J.} \emph{et~al.}
\newblock \bibinfo{title}{Photonics for artificial intelligence and neuromorphic computing}.
\newblock \emph{\bibinfo{journal}{Nature Photonics}} \textbf{\bibinfo{volume}{15}}, \bibinfo{pages}{102--114} (\bibinfo{year}{2021}).

\bibitem{gigan2022imaging}
\bibinfo{author}{Gigan, S.}
\newblock \bibinfo{title}{Imaging and computing with disorder}.
\newblock \emph{\bibinfo{journal}{Nature Physics}} \textbf{\bibinfo{volume}{18}}, \bibinfo{pages}{980--985} (\bibinfo{year}{2022}).

\bibitem{mcmahon2023physics}
\bibinfo{author}{McMahon, P.~L.}
\newblock \bibinfo{title}{The physics of optical computing}.
\newblock \emph{\bibinfo{journal}{Nature Reviews Physics}} \textbf{\bibinfo{volume}{5}}, \bibinfo{pages}{717--734} (\bibinfo{year}{2023}).

\bibitem{vandoorne2008toward}
\bibinfo{author}{Vandoorne, K.} \emph{et~al.}
\newblock \bibinfo{title}{Toward optical signal processing using photonic reservoir computing}.
\newblock \emph{\bibinfo{journal}{Optics Express}} \textbf{\bibinfo{volume}{16}}, \bibinfo{pages}{11182--11192} (\bibinfo{year}{2008}).

\bibitem{paquot2012optoelectronic}
\bibinfo{author}{Paquot, Y.} \emph{et~al.}
\newblock \bibinfo{title}{Optoelectronic reservoir computing}.
\newblock \emph{\bibinfo{journal}{Scientific Reports}} \textbf{\bibinfo{volume}{2}}, \bibinfo{pages}{287} (\bibinfo{year}{2012}).

\bibitem{larger2012photonic}
\bibinfo{author}{Larger, L.} \emph{et~al.}
\newblock \bibinfo{title}{Photonic information processing beyond turing: an optoelectronic implementation of reservoir computing}.
\newblock \emph{\bibinfo{journal}{Optics Express}} \textbf{\bibinfo{volume}{20}}, \bibinfo{pages}{3241--3249} (\bibinfo{year}{2012}).

\bibitem{martinenghi2012photonic}
\bibinfo{author}{Martinenghi, R.}, \bibinfo{author}{Rybalko, S.}, \bibinfo{author}{Jacquot, M.}, \bibinfo{author}{Chembo, Y.~K.} \& \bibinfo{author}{Larger, L.}
\newblock \bibinfo{title}{Photonic nonlinear transient computing with multiple-delay wavelength dynamics}.
\newblock \emph{\bibinfo{journal}{Physical Review Letters}} \textbf{\bibinfo{volume}{108}}, \bibinfo{pages}{244101} (\bibinfo{year}{2012}).

\bibitem{brunner2013parallel}
\bibinfo{author}{Brunner, D.}, \bibinfo{author}{Soriano, M.~C.}, \bibinfo{author}{Mirasso, C.~R.} \& \bibinfo{author}{Fischer, I.}
\newblock \bibinfo{title}{Parallel photonic information processing at gigabyte per second data rates using transient states}.
\newblock \emph{\bibinfo{journal}{Nature Communications}} \textbf{\bibinfo{volume}{4}}, \bibinfo{pages}{1364} (\bibinfo{year}{2013}).

\bibitem{vinckier2015high}
\bibinfo{author}{Vinckier, Q.} \emph{et~al.}
\newblock \bibinfo{title}{High-performance photonic reservoir computer based on a coherently driven passive cavity}.
\newblock \emph{\bibinfo{journal}{Optica}} \textbf{\bibinfo{volume}{2}}, \bibinfo{pages}{438--446} (\bibinfo{year}{2015}).

\bibitem{duport2016fully}
\bibinfo{author}{Duport, F.}, \bibinfo{author}{Smerieri, A.}, \bibinfo{author}{Akrout, A.}, \bibinfo{author}{Haelterman, M.} \& \bibinfo{author}{Massar, S.}
\newblock \bibinfo{title}{Fully analogue photonic reservoir computer}.
\newblock \emph{\bibinfo{journal}{Scientific Reports}} \textbf{\bibinfo{volume}{6}}, \bibinfo{pages}{22381} (\bibinfo{year}{2016}).

\bibitem{larger2017high}
\bibinfo{author}{Larger, L.} \emph{et~al.}
\newblock \bibinfo{title}{High-speed photonic reservoir computing using a time-delay-based architecture: Million words per second classification}.
\newblock \emph{\bibinfo{journal}{Physical Review X}} \textbf{\bibinfo{volume}{7}}, \bibinfo{pages}{011015} (\bibinfo{year}{2017}).

\bibitem{penkovsky2019coupled}
\bibinfo{author}{Penkovsky, B.}, \bibinfo{author}{Porte, X.}, \bibinfo{author}{Jacquot, M.}, \bibinfo{author}{Larger, L.} \& \bibinfo{author}{Brunner, D.}
\newblock \bibinfo{title}{Coupled nonlinear delay systems as deep convolutional neural networks}.
\newblock \emph{\bibinfo{journal}{Physical Review Letters}} \textbf{\bibinfo{volume}{123}}, \bibinfo{pages}{054101} (\bibinfo{year}{2019}).

\bibitem{vandoorne2014experimental}
\bibinfo{author}{Vandoorne, K.} \emph{et~al.}
\newblock \bibinfo{title}{Experimental demonstration of reservoir computing on a silicon photonics chip}.
\newblock \emph{\bibinfo{journal}{Nature Communications}} \textbf{\bibinfo{volume}{5}}, \bibinfo{pages}{3541} (\bibinfo{year}{2014}).

\bibitem{brunner2015reconfigurable}
\bibinfo{author}{Brunner, D.} \& \bibinfo{author}{Fischer, I.}
\newblock \bibinfo{title}{Reconfigurable semiconductor laser networks based on diffractive coupling}.
\newblock \emph{\bibinfo{journal}{Optics Letters}} \textbf{\bibinfo{volume}{40}}, \bibinfo{pages}{3854--3857} (\bibinfo{year}{2015}).

\bibitem{bueno2018reinforcement}
\bibinfo{author}{Bueno, J.} \emph{et~al.}
\newblock \bibinfo{title}{Reinforcement learning in a large-scale photonic recurrent neural network}.
\newblock \emph{\bibinfo{journal}{Optica}} \textbf{\bibinfo{volume}{5}}, \bibinfo{pages}{756--760} (\bibinfo{year}{2018}).

\bibitem{antonik2019human}
\bibinfo{author}{Antonik, P.}, \bibinfo{author}{Marsal, N.}, \bibinfo{author}{Brunner, D.} \& \bibinfo{author}{Rontani, D.}
\newblock \bibinfo{title}{Human action recognition with a large-scale brain-inspired photonic computer}.
\newblock \emph{\bibinfo{journal}{Nature Machine Intelligence}} \textbf{\bibinfo{volume}{1}}, \bibinfo{pages}{530--537} (\bibinfo{year}{2019}).

\bibitem{dong2019optical}
\bibinfo{author}{Dong, J.}, \bibinfo{author}{Rafayelyan, M.}, \bibinfo{author}{Krzakala, F.} \& \bibinfo{author}{Gigan, S.}
\newblock \bibinfo{title}{Optical reservoir computing using multiple light scattering for chaotic systems prediction}.
\newblock \emph{\bibinfo{journal}{IEEE Journal of Selected Topics in Quantum Electronics}} \textbf{\bibinfo{volume}{26}}, \bibinfo{pages}{1--12} (\bibinfo{year}{2019}).

\bibitem{rafayelyan2020large}
\bibinfo{author}{Rafayelyan, M.}, \bibinfo{author}{Dong, J.}, \bibinfo{author}{Tan, Y.}, \bibinfo{author}{Krzakala, F.} \& \bibinfo{author}{Gigan, S.}
\newblock \bibinfo{title}{Large-scale optical reservoir computing for spatiotemporal chaotic systems prediction}.
\newblock \emph{\bibinfo{journal}{Physical Review X}} \textbf{\bibinfo{volume}{10}}, \bibinfo{pages}{041037} (\bibinfo{year}{2020}).

\bibitem{sunada2021photonic}
\bibinfo{author}{Sunada, S.} \& \bibinfo{author}{Uchida, A.}
\newblock \bibinfo{title}{Photonic neural field on a silicon chip: large-scale, high-speed neuro-inspired computing and sensing}.
\newblock \emph{\bibinfo{journal}{Optica}} \textbf{\bibinfo{volume}{8}}, \bibinfo{pages}{1388--1396} (\bibinfo{year}{2021}).

\bibitem{gallicchio2020fast}
\bibinfo{author}{Gallicchio, C.} \& \bibinfo{author}{Micheli, A.}
\newblock \bibinfo{title}{Fast and deep graph neural networks}.
\newblock In \emph{\bibinfo{booktitle}{Proceedings of the AAAI Conference on Artificial Intelligence}}, vol.~\bibinfo{volume}{34}, \bibinfo{pages}{3898--3905} (\bibinfo{year}{2020}).

\bibitem{wang2023echo}
\bibinfo{author}{Wang, S.} \emph{et~al.}
\newblock \bibinfo{title}{Echo state graph neural networks with analogue random resistive memory arrays}.
\newblock \emph{\bibinfo{journal}{Nature Machine Intelligence}} \textbf{\bibinfo{volume}{5}}, \bibinfo{pages}{104--113} (\bibinfo{year}{2023}).

\bibitem{lupo2023deep}
\bibinfo{author}{Lupo, A.}, \bibinfo{author}{Picco, E.}, \bibinfo{author}{Zajnulina, M.} \& \bibinfo{author}{Massar, S.}
\newblock \bibinfo{title}{Deep photonic reservoir computer based on frequency multiplexing with fully analog connection between layers}.
\newblock \emph{\bibinfo{journal}{Optica}} \textbf{\bibinfo{volume}{10}}, \bibinfo{pages}{1478--1485} (\bibinfo{year}{2023}).

\bibitem{shen2023deep}
\bibinfo{author}{Shen, Y.-W.} \emph{et~al.}
\newblock \bibinfo{title}{Deep photonic reservoir computing recurrent network}.
\newblock \emph{\bibinfo{journal}{Optica}} \textbf{\bibinfo{volume}{10}}, \bibinfo{pages}{1745--1751} (\bibinfo{year}{2023}).

\bibitem{gallicchio2017deep}
\bibinfo{author}{Gallicchio, C.}, \bibinfo{author}{Micheli, A.} \& \bibinfo{author}{Pedrelli, L.}
\newblock \bibinfo{title}{Deep reservoir computing: A critical experimental analysis}.
\newblock \emph{\bibinfo{journal}{Neurocomputing}} \textbf{\bibinfo{volume}{268}}, \bibinfo{pages}{87--99} (\bibinfo{year}{2017}).

\bibitem{gauthier2021next}
\bibinfo{author}{Gauthier, D.~J.}, \bibinfo{author}{Bollt, E.}, \bibinfo{author}{Griffith, A.} \& \bibinfo{author}{Barbosa, W.~A.}
\newblock \bibinfo{title}{Next generation reservoir computing}.
\newblock \emph{\bibinfo{journal}{Nature Communications}} \textbf{\bibinfo{volume}{12}}, \bibinfo{pages}{5564} (\bibinfo{year}{2021}).

\bibitem{pyle2021domain}
\bibinfo{author}{Pyle, R.}, \bibinfo{author}{Jovanovic, N.}, \bibinfo{author}{Subramanian, D.}, \bibinfo{author}{Palem, K.~V.} \& \bibinfo{author}{Patel, A.~B.}
\newblock \bibinfo{title}{Domain-driven models yield better predictions at lower cost than reservoir computers in lorenz systems}.
\newblock \emph{\bibinfo{journal}{Philosophical Transactions of the Royal Society A}} \textbf{\bibinfo{volume}{379}}, \bibinfo{pages}{20200246} (\bibinfo{year}{2021}).

\bibitem{Jauriguee23121560}
\bibinfo{author}{Jaurigue, L.}, \bibinfo{author}{Robertson, E.}, \bibinfo{author}{Wolters, J.} \& \bibinfo{author}{Lüdge, K.}
\newblock \bibinfo{title}{Reservoir computing with delayed input for fast and easy optimisation}.
\newblock \emph{\bibinfo{journal}{Entropy}} \textbf{\bibinfo{volume}{23}} (\bibinfo{year}{2021}).

\bibitem{Jaurigue_2024}
\bibinfo{author}{Jaurigue, L.} \& \bibinfo{author}{Lüdge, K.}
\newblock \bibinfo{title}{Reducing reservoir computer hyperparameter dependence by external timescale tailoring}.
\newblock \emph{\bibinfo{journal}{Neuromorphic Computing and Engineering}} \textbf{\bibinfo{volume}{4}}, \bibinfo{pages}{014001} (\bibinfo{year}{2024}).

\bibitem{pathak2017using}
\bibinfo{author}{Pathak, J.}, \bibinfo{author}{Lu, Z.}, \bibinfo{author}{Hunt, B.~R.}, \bibinfo{author}{Girvan, M.} \& \bibinfo{author}{Ott, E.}
\newblock \bibinfo{title}{Using machine learning to replicate chaotic attractors and calculate lyapunov exponents from data}.
\newblock \emph{\bibinfo{journal}{Chaos: An Interdisciplinary Journal of Nonlinear Science}} \textbf{\bibinfo{volume}{27}} (\bibinfo{year}{2017}).

\bibitem{lu2018attractor}
\bibinfo{author}{Lu, Z.}, \bibinfo{author}{Hunt, B.~R.} \& \bibinfo{author}{Ott, E.}
\newblock \bibinfo{title}{Attractor reconstruction by machine learning}.
\newblock \emph{\bibinfo{journal}{Chaos: An Interdisciplinary Journal of Nonlinear Science}} \textbf{\bibinfo{volume}{28}} (\bibinfo{year}{2018}).

\bibitem{griffith2019forecasting}
\bibinfo{author}{Griffith, A.}, \bibinfo{author}{Pomerance, A.} \& \bibinfo{author}{Gauthier, D.~J.}
\newblock \bibinfo{title}{Forecasting chaotic systems with very low connectivity reservoir computers}.
\newblock \emph{\bibinfo{journal}{Chaos: An Interdisciplinary Journal of Nonlinear Science}} \textbf{\bibinfo{volume}{29}} (\bibinfo{year}{2019}).

\bibitem{pedrelli2021hierarchical}
\bibinfo{author}{Pedrelli, L.} \& \bibinfo{author}{Hinaut, X.}
\newblock \bibinfo{title}{Hierarchical-task reservoir for online semantic analysis from continuous speech}.
\newblock \emph{\bibinfo{journal}{IEEE Transactions on Neural Networks and Learning Systems}} \textbf{\bibinfo{volume}{33}}, \bibinfo{pages}{2654--2663} (\bibinfo{year}{2021}).

\bibitem{barbosa2022learning}
\bibinfo{author}{Barbosa, W.~A.} \& \bibinfo{author}{Gauthier, D.~J.}
\newblock \bibinfo{title}{Learning spatiotemporal chaos using next-generation reservoir computing}.
\newblock \emph{\bibinfo{journal}{Chaos: An Interdisciplinary Journal of Nonlinear Science}} \textbf{\bibinfo{volume}{32}} (\bibinfo{year}{2022}).

\bibitem{kim2023neural}
\bibinfo{author}{Kim, J.~Z.} \& \bibinfo{author}{Bassett, D.~S.}
\newblock \bibinfo{title}{A neural machine code and programming framework for the reservoir computer}.
\newblock \emph{\bibinfo{journal}{Nature Machine Intelligence}} \textbf{\bibinfo{volume}{5}}, \bibinfo{pages}{622--630} (\bibinfo{year}{2023}).

\bibitem{delloye2023optical}
\bibinfo{author}{Delloye, L.} \emph{et~al.}
\newblock \bibinfo{title}{An optical ising spin glass simulator with tuneable short range couplings}.
\newblock \emph{\bibinfo{journal}{arXiv preprint arXiv:2309.10764}}  (\bibinfo{year}{2023}).

\bibitem{ohana2020kernel}
\bibinfo{author}{Ohana, R.} \emph{et~al.}
\newblock \bibinfo{title}{Kernel computations from large-scale random features obtained by optical processing units}.
\newblock In \emph{\bibinfo{booktitle}{ICASSP 2020-2020 IEEE International Conference on Acoustics, Speech and Signal Processing (ICASSP)}}, \bibinfo{pages}{9294--9298} (\bibinfo{organization}{IEEE}, \bibinfo{year}{2020}).

\bibitem{pierangeli2021scalable}
\bibinfo{author}{Pierangeli, D.}, \bibinfo{author}{Rafayelyan, M.}, \bibinfo{author}{Conti, C.} \& \bibinfo{author}{Gigan, S.}
\newblock \bibinfo{title}{Scalable spin-glass optical simulator}.
\newblock \emph{\bibinfo{journal}{Physical Review Applied}} \textbf{\bibinfo{volume}{15}}, \bibinfo{pages}{034087} (\bibinfo{year}{2021}).

\bibitem{kent2024controlling}
\bibinfo{author}{Kent, R.~M.}, \bibinfo{author}{Barbosa, W.~A.} \& \bibinfo{author}{Gauthier, D.~J.}
\newblock \bibinfo{title}{Controlling chaos using edge computing hardware}.
\newblock \emph{\bibinfo{journal}{Nature Communications}} \textbf{\bibinfo{volume}{15}}, \bibinfo{pages}{3886} (\bibinfo{year}{2024}).

\bibitem{cox2024photonic}
\bibinfo{author}{Cox, N.}, \bibinfo{author}{Murray, J.}, \bibinfo{author}{Hart, J.} \& \bibinfo{author}{Redding, B.}
\newblock \bibinfo{title}{Photonic next-generation reservoir computer based on distributed feedback in optical fiber}.
\newblock \emph{\bibinfo{journal}{arXiv:2404.07116}}  (\bibinfo{year}{2024}).

\bibitem{lorenz1963deterministic}
\bibinfo{author}{Lorenz, E.~N.}
\newblock \bibinfo{title}{Deterministic nonperiodic flow}.
\newblock \emph{\bibinfo{journal}{Journal of atmospheric sciences}} \textbf{\bibinfo{volume}{20}}, \bibinfo{pages}{130--141} (\bibinfo{year}{1963}).

\bibitem{hohenberg1989chaotic}
\bibinfo{author}{Hohenberg, P.} \& \bibinfo{author}{Shraiman, B.~I.}
\newblock \bibinfo{title}{Chaotic behavior of an extended system}.
\newblock \emph{\bibinfo{journal}{Physica D: Nonlinear Phenomena}} \textbf{\bibinfo{volume}{37}}, \bibinfo{pages}{109--115} (\bibinfo{year}{1989}).

\bibitem{edson2019lyapunov}
\bibinfo{author}{Edson, R.~A.}, \bibinfo{author}{Bunder, J.~E.}, \bibinfo{author}{Mattner, T.~W.} \& \bibinfo{author}{Roberts, A.~J.}
\newblock \bibinfo{title}{Lyapunov exponents of the kuramoto--sivashinsky pde}.
\newblock \emph{\bibinfo{journal}{The ANZIAM Journal}} \textbf{\bibinfo{volume}{61}}, \bibinfo{pages}{270--285} (\bibinfo{year}{2019}).

\bibitem{dong2020reservoir}
\bibinfo{author}{Dong, J.}, \bibinfo{author}{Ohana, R.}, \bibinfo{author}{Rafayelyan, M.} \& \bibinfo{author}{Krzakala, F.}
\newblock \bibinfo{title}{Reservoir computing meets recurrent kernels and structured transforms}.
\newblock \emph{\bibinfo{journal}{Advances in Neural Information Processing Systems}} \textbf{\bibinfo{volume}{33}}, \bibinfo{pages}{16785--16796} (\bibinfo{year}{2020}).

\bibitem{yperman2016bayesian}
\bibinfo{author}{Yperman, J.} \& \bibinfo{author}{Becker, T.}
\newblock \bibinfo{title}{Bayesian optimization of hyper-parameters in reservoir computing}.
\newblock \emph{\bibinfo{journal}{arXiv:1611.05193}}  (\bibinfo{year}{2016}).

\bibitem{shahriari2015taking}
\bibinfo{author}{Shahriari, B.}, \bibinfo{author}{Swersky, K.}, \bibinfo{author}{Wang, Z.}, \bibinfo{author}{Adams, R.~P.} \& \bibinfo{author}{De~Freitas, N.}
\newblock \bibinfo{title}{Taking the human out of the loop: A review of {B}ayesian optimization}.
\newblock \emph{\bibinfo{journal}{Proceedings of the IEEE}} \textbf{\bibinfo{volume}{104}}, \bibinfo{pages}{148--175} (\bibinfo{year}{2015}).

\end{thebibliography}


\begin{thebibliography}{10}
\expandafter\ifx\csname url\endcsname\relax
  \def\url#1{\texttt{#1}}\fi
\expandafter\ifx\csname urlprefix\endcsname\relax\def\urlprefix{URL }\fi
\providecommand{\bibinfo}[2]{#2}
\providecommand{\eprint}[2][]{\url{#2}}

\bibitem{gigan2022imaging}
\bibinfo{author}{Gigan, S.}
\newblock \bibinfo{title}{Imaging and computing with disorder}.
\newblock \emph{\bibinfo{journal}{Nature Physics}} \textbf{\bibinfo{volume}{18}}, \bibinfo{pages}{980--985} (\bibinfo{year}{2022}).

\bibitem{bingham2001random}
\bibinfo{author}{Bingham, E.} \& \bibinfo{author}{Mannila, H.}
\newblock \bibinfo{title}{Random projection in dimensionality reduction: applications to image and text data}.
\newblock In \emph{\bibinfo{booktitle}{Proceedings of the seventh ACM SIGKDD international conference on Knowledge discovery and data mining}}, \bibinfo{pages}{245--250} (\bibinfo{year}{2001}).

\bibitem{dong2019optical}
\bibinfo{author}{Dong, J.}, \bibinfo{author}{Rafayelyan, M.}, \bibinfo{author}{Krzakala, F.} \& \bibinfo{author}{Gigan, S.}
\newblock \bibinfo{title}{Optical reservoir computing using multiple light scattering for chaotic systems prediction}.
\newblock \emph{\bibinfo{journal}{IEEE Journal of Selected Topics in Quantum Electronics}} \textbf{\bibinfo{volume}{26}}, \bibinfo{pages}{1--12} (\bibinfo{year}{2019}).

\bibitem{rafayelyan2020large}
\bibinfo{author}{Rafayelyan, M.}, \bibinfo{author}{Dong, J.}, \bibinfo{author}{Tan, Y.}, \bibinfo{author}{Krzakala, F.} \& \bibinfo{author}{Gigan, S.}
\newblock \bibinfo{title}{Large-scale optical reservoir computing for spatiotemporal chaotic systems prediction}.
\newblock \emph{\bibinfo{journal}{Physical Review X}} \textbf{\bibinfo{volume}{10}}, \bibinfo{pages}{041037} (\bibinfo{year}{2020}).

\bibitem{saade2016random}
\bibinfo{author}{Saade, A.} \emph{et~al.}
\newblock \bibinfo{title}{Random projections through multiple optical scattering: Approximating kernels at the speed of light}.
\newblock In \emph{\bibinfo{booktitle}{2016 IEEE International Conference on Acoustics, Speech and Signal Processing (ICASSP)}}, \bibinfo{pages}{6215--6219} (\bibinfo{organization}{IEEE}, \bibinfo{year}{2016}).

\bibitem{leonetti2021optical}
\bibinfo{author}{Leonetti, M.}, \bibinfo{author}{H{\"o}rmann, E.}, \bibinfo{author}{Leuzzi, L.}, \bibinfo{author}{Parisi, G.} \& \bibinfo{author}{Ruocco, G.}
\newblock \bibinfo{title}{Optical computation of a spin glass dynamics with tunable complexity}.
\newblock \emph{\bibinfo{journal}{Proceedings of the National Academy of Sciences}} \textbf{\bibinfo{volume}{118}}, \bibinfo{pages}{e2015207118} (\bibinfo{year}{2021}).

\bibitem{pierangeli2021scalable}
\bibinfo{author}{Pierangeli, D.}, \bibinfo{author}{Rafayelyan, M.}, \bibinfo{author}{Conti, C.} \& \bibinfo{author}{Gigan, S.}
\newblock \bibinfo{title}{Scalable spin-glass optical simulator}.
\newblock \emph{\bibinfo{journal}{Physical Review Applied}} \textbf{\bibinfo{volume}{15}}, \bibinfo{pages}{034087} (\bibinfo{year}{2021}).

\bibitem{matthes2019optical}
\bibinfo{author}{Matth{\`e}s, M.~W.}, \bibinfo{author}{Del~Hougne, P.}, \bibinfo{author}{De~Rosny, J.}, \bibinfo{author}{Lerosey, G.} \& \bibinfo{author}{Popoff, S.~M.}
\newblock \bibinfo{title}{Optical complex media as universal reconfigurable linear operators}.
\newblock \emph{\bibinfo{journal}{Optica}} \textbf{\bibinfo{volume}{6}}, \bibinfo{pages}{465--472} (\bibinfo{year}{2019}).

\bibitem{launay2020hardware}
\bibinfo{author}{Launay, J.} \emph{et~al.}
\newblock \bibinfo{title}{Hardware beyond backpropagation: a photonic co-processor for direct feedback alignment}.
\newblock \emph{\bibinfo{journal}{arXiv:2012.06373}}  (\bibinfo{year}{2020}).

\bibitem{ghanem2021fast}
\bibinfo{author}{Ghanem, H.}, \bibinfo{author}{Keriven, N.} \& \bibinfo{author}{Tremblay, N.}
\newblock \bibinfo{title}{Fast graph kernel with optical random features}.
\newblock In \emph{\bibinfo{booktitle}{ICASSP 2021-2021 IEEE International Conference on Acoustics, Speech and Signal Processing (ICASSP)}}, \bibinfo{pages}{3575--3579} (\bibinfo{organization}{IEEE}, \bibinfo{year}{2021}).

\bibitem{keriven2020newma}
\bibinfo{author}{Keriven, N.}, \bibinfo{author}{Garreau, D.} \& \bibinfo{author}{Poli, I.}
\newblock \bibinfo{title}{Newma: a new method for scalable model-free online change-point detection}.
\newblock \emph{\bibinfo{journal}{IEEE Transactions on Signal Processing}} \textbf{\bibinfo{volume}{68}}, \bibinfo{pages}{3515--3528} (\bibinfo{year}{2020}).

\bibitem{gauthier2021next}
\bibinfo{author}{Gauthier, D.~J.}, \bibinfo{author}{Bollt, E.}, \bibinfo{author}{Griffith, A.} \& \bibinfo{author}{Barbosa, W.~A.}
\newblock \bibinfo{title}{Next generation reservoir computing}.
\newblock \emph{\bibinfo{journal}{Nature Communications}} \textbf{\bibinfo{volume}{12}}, \bibinfo{pages}{5564} (\bibinfo{year}{2021}).

\bibitem{estebanez2019constructive}
\bibinfo{author}{Est{\'e}banez, I.}, \bibinfo{author}{Fischer, I.} \& \bibinfo{author}{Soriano, M.~C.}
\newblock \bibinfo{title}{Constructive role of noise for high-quality replication of chaotic attractor dynamics using a hardware-based reservoir computer}.
\newblock \emph{\bibinfo{journal}{Physical Review Applied}} \textbf{\bibinfo{volume}{12}}, \bibinfo{pages}{034058} (\bibinfo{year}{2019}).

\bibitem{zhou2021large}
\bibinfo{author}{Zhou, T.} \emph{et~al.}
\newblock \bibinfo{title}{Large-scale neuromorphic optoelectronic computing with a reconfigurable diffractive processing unit}.
\newblock \emph{\bibinfo{journal}{Nature Photonics}} \textbf{\bibinfo{volume}{15}}, \bibinfo{pages}{367--373} (\bibinfo{year}{2021}).

\bibitem{tzang2019wavefront}
\bibinfo{author}{Tzang, O.} \emph{et~al.}
\newblock \bibinfo{title}{Wavefront shaping in complex media with a 350 khz modulator via a 1d-to-2d transform}.
\newblock \emph{\bibinfo{journal}{Nature Photonics}} \textbf{\bibinfo{volume}{13}}, \bibinfo{pages}{788--793} (\bibinfo{year}{2019}).

\bibitem{trajtenberg2024lnos}
\bibinfo{author}{Trajtenberg-Mills, S.} \emph{et~al.}
\newblock \bibinfo{title}{Lnos: Lithium niobate on silicon spatial light modulator}.
\newblock \emph{\bibinfo{journal}{arXiv preprint arXiv:2402.14608}}  (\bibinfo{year}{2024}).

\bibitem{rogers2021universal}
\bibinfo{author}{Rogers, C.} \emph{et~al.}
\newblock \bibinfo{title}{A universal 3d imaging sensor on a silicon photonics platform}.
\newblock \emph{\bibinfo{journal}{Nature}} \textbf{\bibinfo{volume}{590}}, \bibinfo{pages}{256--261} (\bibinfo{year}{2021}).

\bibitem{morimoto2020megapixel}
\bibinfo{author}{Morimoto, K.} \emph{et~al.}
\newblock \bibinfo{title}{Megapixel time-gated spad image sensor for 2d and 3d imaging applications}.
\newblock \emph{\bibinfo{journal}{Optica}} \textbf{\bibinfo{volume}{7}}, \bibinfo{pages}{346--354} (\bibinfo{year}{2020}).

\bibitem{li2022fpga}
\bibinfo{author}{Li, Y.}, \bibinfo{author}{Li, S.~E.}, \bibinfo{author}{Jia, X.}, \bibinfo{author}{Zeng, S.} \& \bibinfo{author}{Wang, Y.}
\newblock \bibinfo{title}{{FPGA} accelerated model predictive control for autonomous driving}.
\newblock \emph{\bibinfo{journal}{Journal of intelligent and connected vehicles}} \textbf{\bibinfo{volume}{5}}, \bibinfo{pages}{63--71} (\bibinfo{year}{2022}).

\bibitem{khairy2020tpu}
\bibinfo{author}{Khairy, M.}
\newblock \bibinfo{title}{Tpu vs gpu vs cerebras vs graphcore: a fair comparison between ml hardware} (\bibinfo{year}{2020}).

\bibitem{ohana2020kernel}
\bibinfo{author}{Ohana, R.} \emph{et~al.}
\newblock \bibinfo{title}{Kernel computations from large-scale random features obtained by optical processing units}.
\newblock In \emph{\bibinfo{booktitle}{ICASSP 2020-2020 IEEE International Conference on Acoustics, Speech and Signal Processing (ICASSP)}}, \bibinfo{pages}{9294--9298} (\bibinfo{organization}{IEEE}, \bibinfo{year}{2020}).

\bibitem{barbosa2022learning}
\bibinfo{author}{Barbosa, W.~A.} \& \bibinfo{author}{Gauthier, D.~J.}
\newblock \bibinfo{title}{Learning spatiotemporal chaos using next-generation reservoir computing}.
\newblock \emph{\bibinfo{journal}{Chaos: An Interdisciplinary Journal of Nonlinear Science}} \textbf{\bibinfo{volume}{32}} (\bibinfo{year}{2022}).

\bibitem{cai2018feature}
\bibinfo{author}{Cai, J.}, \bibinfo{author}{Luo, J.}, \bibinfo{author}{Wang, S.} \& \bibinfo{author}{Yang, S.}
\newblock \bibinfo{title}{Feature selection in machine learning: A new perspective}.
\newblock \emph{\bibinfo{journal}{Neurocomputing}} \textbf{\bibinfo{volume}{300}}, \bibinfo{pages}{70--79} (\bibinfo{year}{2018}).

\bibitem{pathak2017using}
\bibinfo{author}{Pathak, J.}, \bibinfo{author}{Lu, Z.}, \bibinfo{author}{Hunt, B.~R.}, \bibinfo{author}{Girvan, M.} \& \bibinfo{author}{Ott, E.}
\newblock \bibinfo{title}{Using machine learning to replicate chaotic attractors and calculate lyapunov exponents from data}.
\newblock \emph{\bibinfo{journal}{Chaos: An Interdisciplinary Journal of Nonlinear Science}} \textbf{\bibinfo{volume}{27}} (\bibinfo{year}{2017}).

\bibitem{nakajima2021reservoir}
\bibinfo{author}{Nakajima, K.} \& \bibinfo{author}{Fischer, I.}
\newblock \emph{\bibinfo{title}{Reservoir Computing}} (\bibinfo{publisher}{Springer}, \bibinfo{year}{2021}).

\bibitem{pathak2018model}
\bibinfo{author}{Pathak, J.}, \bibinfo{author}{Hunt, B.}, \bibinfo{author}{Girvan, M.}, \bibinfo{author}{Lu, Z.} \& \bibinfo{author}{Ott, E.}
\newblock \bibinfo{title}{Model-free prediction of large spatiotemporally chaotic systems from data: A reservoir computing approach}.
\newblock \emph{\bibinfo{journal}{Physical Review Letters}} \textbf{\bibinfo{volume}{120}}, \bibinfo{pages}{024102} (\bibinfo{year}{2018}).

\bibitem{vlachas2020backpropagation}
\bibinfo{author}{Vlachas, P.-R.} \emph{et~al.}
\newblock \bibinfo{title}{Backpropagation algorithms and reservoir computing in recurrent neural networks for the forecasting of complex spatiotemporal dynamics}.
\newblock \emph{\bibinfo{journal}{Neural Networks}} \textbf{\bibinfo{volume}{126}}, \bibinfo{pages}{191--217} (\bibinfo{year}{2020}).

\bibitem{lu2018attractor}
\bibinfo{author}{Lu, Z.}, \bibinfo{author}{Hunt, B.~R.} \& \bibinfo{author}{Ott, E.}
\newblock \bibinfo{title}{Attractor reconstruction by machine learning}.
\newblock \emph{\bibinfo{journal}{Chaos: An Interdisciplinary Journal of Nonlinear Science}} \textbf{\bibinfo{volume}{28}} (\bibinfo{year}{2018}).

\bibitem{wikner2021using}
\bibinfo{author}{Wikner, A.} \emph{et~al.}
\newblock \bibinfo{title}{Using data assimilation to train a hybrid forecast system that combines machine-learning and knowledge-based components}.
\newblock \emph{\bibinfo{journal}{Chaos: An Interdisciplinary Journal of Nonlinear Science}} \textbf{\bibinfo{volume}{31}} (\bibinfo{year}{2021}).

\bibitem{jiang2019model}
\bibinfo{author}{Jiang, J.} \& \bibinfo{author}{Lai, Y.-C.}
\newblock \bibinfo{title}{Model-free prediction of spatiotemporal dynamical systems with recurrent neural networks: Role of network spectral radius}.
\newblock \emph{\bibinfo{journal}{Physical review research}} \textbf{\bibinfo{volume}{1}}, \bibinfo{pages}{033056} (\bibinfo{year}{2019}).

\bibitem{dong2020reservoir}
\bibinfo{author}{Dong, J.}, \bibinfo{author}{Ohana, R.}, \bibinfo{author}{Rafayelyan, M.} \& \bibinfo{author}{Krzakala, F.}
\newblock \bibinfo{title}{Reservoir computing meets recurrent kernels and structured transforms}.
\newblock \emph{\bibinfo{journal}{Advances in Neural Information Processing Systems}} \textbf{\bibinfo{volume}{33}}, \bibinfo{pages}{16785--16796} (\bibinfo{year}{2020}).

\bibitem{PhysRevApplied.7.054014}
\bibinfo{author}{Antonik, P.}, \bibinfo{author}{Haelterman, M.} \& \bibinfo{author}{Massar, S.}
\newblock \bibinfo{title}{Brain-inspired photonic signal processor for generating periodic patterns and emulating chaotic systems}.
\newblock \emph{\bibinfo{journal}{Phys. Rev. Appl.}} \textbf{\bibinfo{volume}{7}}, \bibinfo{pages}{054014} (\bibinfo{year}{2017}).

\end{thebibliography}

\end{document}

\title{Supplementary information for:\\ Optical next generation reservoir computing}

\author{Hao Wang$^{1,2,\ast}$, 
        Jianqi Hu$^{1,4,\ast,\dagger}$, 
        YoonSeok Baek$^{1}$,       
        Kohei Tsuchiyama$^{1,3}$,
        Malo Joly$^{1}$,
        Qiang Liu$^{2,\dagger}$,  
        and Sylvain Gigan$^{1,\dagger}$}
\affiliation{
$^1$Laboratoire Kastler Brossel\char`,{} École Normale Supérieure - Paris Sciences et Lettres (PSL) Research University\char`,{} Sorbonne Université\char`,{} Centre National de la Recherche Scientifique (CNRS)\char`,{} UMR 8552\char`,{} Collège de France\char`,{} 24 rue Lhomond\char`,{} 75005 Paris\char`,{} France.\\
$^2$State Key Laboratory of Precision Space-time Information Sensing Technology\char`,{} Department of Precision Instrument\char`,{} Tsinghua University\char`,{} Beijing 100084\char`,{} China.\\
$^3$Department of Information Physics and Computing\char`,{} Graduate School of Information Science and Technology\char`,{} The University of Tokyo\char`,{} 7-3-1 Hongo\char`,{} Bunkyo-ku\char`,{} Tokyo 113-8656\char`,{} Japan.\\
$^4$Present address: Swiss Federal Institute of Technology Lausanne (EPFL)\char`,{} CH-1015 Lausanne\char`,{} Switzerland.\\
}

\maketitle

\noindent{\textbf{\large{Contents}}}
\vspace{0.1cm}

\noindent{\textbf{Supplementary Notes:}}\\
\noindent{1. The principle of optical NGRC}\\
2. Experimental setup details\\
3. Simulation comparison of optical NGRC and optical conventional RC\\
4. The impact of device quantization on optical NGRC\\
5. Optical computation analysis

\vspace{0.1cm}
\noindent{\textbf{Supplementary Figures:}}\\
\noindent{Figure S1. Optical NGRC experimental system}\\
Figure S2. Optical NGRC principle\\
Figure S3. Illustration of optical NGRC formulation in matrix representation.\\
Figure S4. Comparison of the effective readout matrices of optical NGRC and digital NGRC in the same feature basis.\\
Figure S5. Experimental system stability\\
Figure S6. Bayesian optimization log during the short-term prediction of KS time series experiments\\
Figure S7. Performance comparison of optical NGRC and optical conventional RC based on scattering media in simulation\\
Figure S8. Simulation of optical NGRC forecasting errors based on different quantization bit depths of devices

\vspace{0.1cm}
\noindent{\textbf{Supplementary Algorithms:}}\\
\noindent{Algorithm S1. Optical NGRC for forecasting dynamical systems}\\
Algorithm S2. Optical NGRC for deducing unmeasured variables of dynamical systems

\vspace{0.1cm}
\noindent{\textbf{Supplementary Tables:}}\\
\noindent{Table S1. Comparison of optical NGRC, optical conventional RC and digital NGRC}\\
Table S2. Summary of data encoding and processing parameters used in the experiments \\
Table S3. Performance comparison with previous works on Lorenz63 and KS time-series prediction

%%%%%%%%%%%%%%%%%%%%%
\section*{Supplementary Note 1. The principle of optical NGRC} 
Multiple scattering phenomenon in optics has recently been harnessed as a computational resource owing to its inherent complexity and high dimensionality \cite{gigan2022imaging}. As coherent laser light travels through a disordered medium, it experiences random scattering and numerous interference events occur. This process results in the formation of a speckle field at the output. For a given fixed scattering medium, the input and output optical fields are deterministically related by a complex matrix $\boldsymbol{W}$, known as the transmission matrix (TM), i.e., $\boldsymbol{E}_{out} = \boldsymbol{W} \cdot \boldsymbol{E}_{in}$. Therefore, despite the apparent randomness of the speckle field, it encapsulates rich information of the input as speckle features. Experiments and theoretical studies reveal that the real and imaginary parts of the entries in the TM follow Gaussian independent and identical distributions (see Supplementary Figs. \ref{FigureS1}b-c). This spurs the recent research interests in harnessing such a  disordered optical process for signal processing tasks \cite{gigan2022imaging}. In essence, this process can be conceptualized as a matrix-vector multiplication, where the input data $\boldsymbol{E}_{in}$ is multiplied by a random matrix $\boldsymbol{W}$. Au such, it performs optically the random projection, a ubiquitous mathematical operation used in many signal processing scenarios \cite{bingham2001random}. The optical setup with multiple light scattering executes random projection in a fast way without the need to measure or digitally store the random matrix, which can reach an extreme scale where the benefits of optical computing become pronounced. Following this spirit, optical random projection has been successfully applied in diverse computing and signal processing tasks including reservoir computing \cite{dong2019optical,rafayelyan2020large}, extreme learning \cite{saade2016random}, spin-glass simulator \cite{leonetti2021optical,pierangeli2021scalable}, reconfigurable linear operators \cite{matthes2019optical}, direct feedback alignment training \cite{launay2020hardware}, graph kernel \cite{ghanem2021fast}, online change-point detection \cite{keriven2020newma}, among others.

\begin{figure*}[!htb]
  \renewcommand{\figurename}{Supplementary Figure}
  \centering{
  \includegraphics[width = 1.0\linewidth]{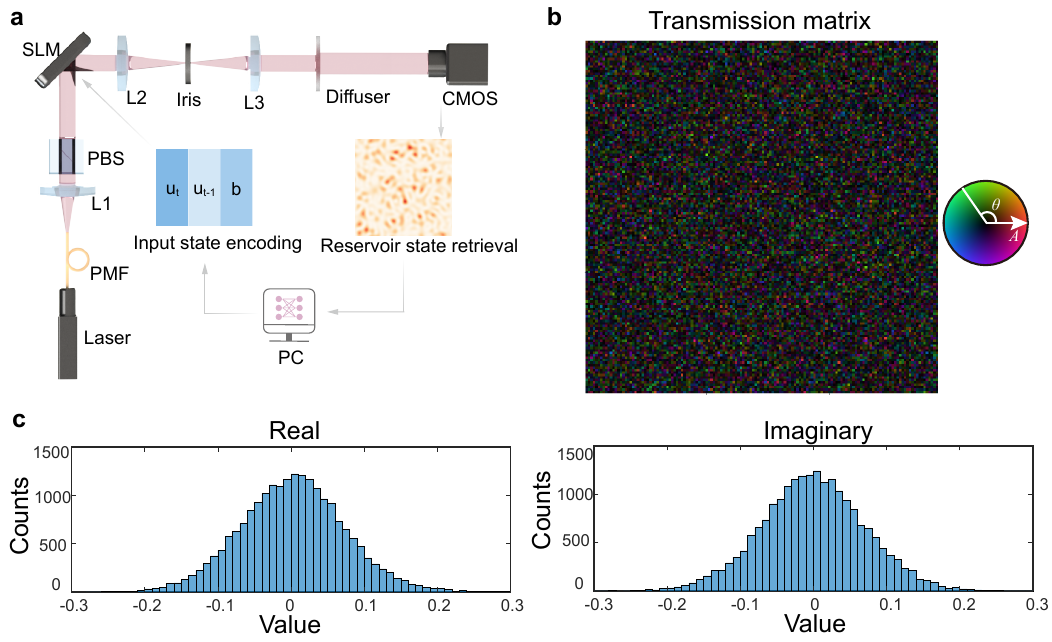}
  } 
  \caption{\noindent\textbf{Optical NGRC experimental system.} \textbf{a} Experimental setup. PMF: polarization-maintaining fiber; PBS: polarizing beam splitter; SLM: spatial light modulator; Diffuser: ground glass diffuser; CMOS: camera; L1, L2, L3: lens; PC: personal computer. \textbf{b} Typical experimental transmission matrix of the current setup ($144\times144$). \textbf{c} Statistical distribution of the entries of the TM in \textbf{b} (left panel: real components; right panel: imaginary components).}
 \label{FigureS1}
\end{figure*} 

In our optical system, we employ a phase-only SLM to encode input data $\boldsymbol{x}=[x_1, x_2, ..., x_M]^T$, a scattering medium to perform random mixing (feature extraction), and a camera to retrieve the resulting speckle intensity features (reservoir features) $\boldsymbol{y}=[y_1,y_2, ...,y_N]^T$. Although light scattering itself is fundamentally linear in this work, our system achieves a nonlinear mapping between the input data and the output (reservoir) features because of the strategic incorporation of phase encoding and intensity detection nonlinearities, as described by $\boldsymbol{y}=|\boldsymbol{W} \cdot {\rm exp}(i\boldsymbol{x})|^2$. Therefore, the combination of linear random projection and element-wise nonlinearity provides rich high order nonlinear terms, which can be used for the construction of NGRC. Collectively, we can rewrite the output intensity at the $n$-th output mode using Taylor expansion:
\begin{equation}
    y^{(n)} \approx \alpha^{(n)}_{00} + \underbrace{\alpha^{(n)}_{10} x_1 + \alpha^{(n)}_{20} x_2 +...}_{\mathrm{Linear \thinspace terms \thinspace}} + \underbrace{\alpha^{(n)}_{11} x_1^2 + ...+ \alpha^{(n)}_{12} x_1 x_2 + ...}_{\mathrm{Quadratic \thinspace terms}} + ...,
    \label{eq_s1}
\end{equation}
where $\alpha^{(n)}_{ij}$ represents the weighted coefficient (here the nomenclature of the subscript is shown up to quadratic terms for clarity). The equation above implies that the optical nonlinear mapping of our setup is equivalent to firstly calculating the rich monomials (polynomial terms) of the input data explicitly, and then linearly combining them into speckle intensity features. This understanding is crucial to link optical NGRC in this work and digital NGRC \cite{gauthier2021next}. Experimentally, by encoding multiple time steps of input time series data into $\boldsymbol{x}$ (e.g., $[\boldsymbol{u}_t, \boldsymbol{u}_{t-1}]^T \rightarrow{\boldsymbol{x}}$), and specifying the reservoir state $\boldsymbol{r}_{t+1}$ at the time step $t+1$ from $\boldsymbol{y}$ in Eq. \eqref{eq_s1} ($\boldsymbol{y}\rightarrow \boldsymbol{r}_{t+1}$) as well as grouping all the coefficients $\alpha^{(n)}_{ij}$ into $\boldsymbol{M}_s$, we derive Eq. (4) in the main text. The explicit polynomial feature terms of the input data are compiled into a feature vector denoted by $\boldsymbol{\Theta_t}$. In other words, $\boldsymbol{\Theta_t}$ is the reservoir feature terms at the time step $t$ of NGRC, namely $\boldsymbol{\Theta_t} = (1, \boldsymbol{u}_t^T,\boldsymbol{u}_{t-1}^T,
    \mathbb{U}(\boldsymbol{u}_{t} \otimes \boldsymbol{u}_{t}), \mathbb{U}(\boldsymbol{u}_{t-1} \otimes \boldsymbol{u}_{t-1}), \mathbb{U}(\boldsymbol{u}_{t} \otimes \boldsymbol{u}_{t-1}), ...)$ with $\mathbb{U}$ as an operation to collect all unique monomials from the matrix vectorization of the outer product of two vectors. And the system-given matrix $\boldsymbol{M}_s$ incorporates the phase encoding, TM, and intensity detection of the optical setup altogether. Upon formulating the reservoir state as $\boldsymbol{r}_{t+1}\approx \boldsymbol{M}_s\cdot \boldsymbol{\Theta_t}$, we optimize a linear digital readout layer $\boldsymbol{W}_{out}$ atop the reservoir states for prediction, i.e., $\boldsymbol{o}_{t}=\boldsymbol{W}_{out} \boldsymbol{r}_{t} \approx \boldsymbol{W}_{out} \boldsymbol{M}_s \boldsymbol{\Theta_t}$. This is equivalent to directly harnessing the polynomial feature terms $\boldsymbol{\Theta_t}$ by $\boldsymbol{W}_{out}^{\prime}  \boldsymbol{M}_s$, which is at the heart of NGRC. We visualize the working principle of optical NGRC in Supplementary Fig. \ref{FigureS2}.

\begin{figure*}[!htb]
  \renewcommand{\figurename}{Supplementary Figure}
  \centering{
  \includegraphics[width = 0.8\linewidth]{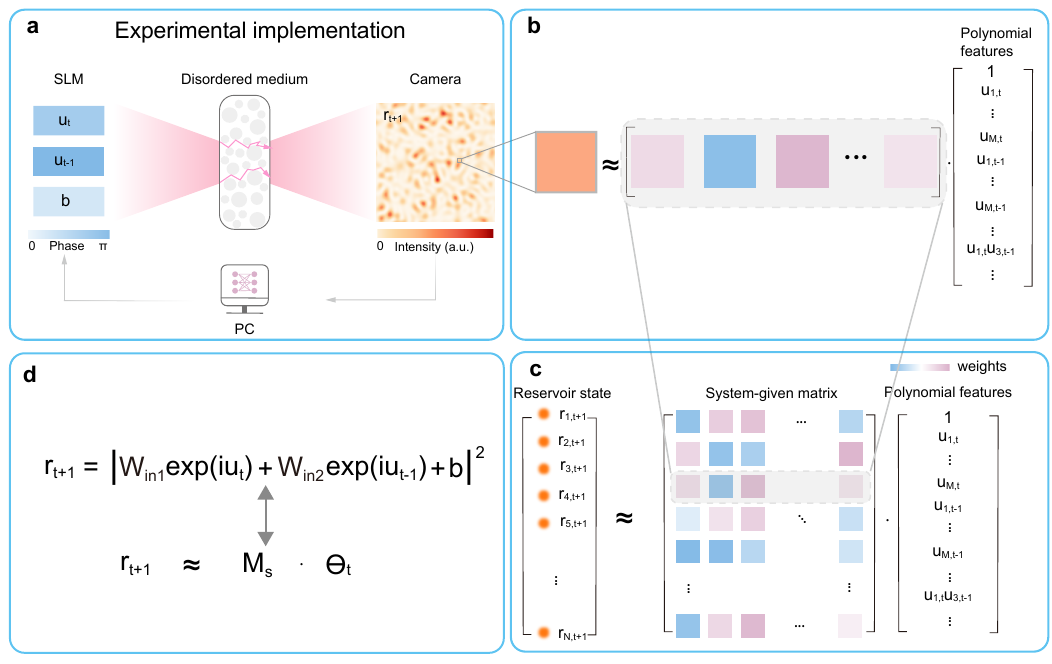}
  } 
  \caption{\noindent\textbf{Optical NGRC principle.} \textbf{a} Optical implementation. \textbf{b} Decomposition of the speckle intensity feature at an output mode. \textbf{c} Mathematical model of optical NGRC. \textbf{d} Matrix representation of optical NGRC.}
 \label{FigureS2}
\end{figure*} 

\begin{figure*}[!htb]
  \renewcommand{\figurename}{Supplementary Figure}
  \centering{
  \includegraphics[width = 0.8\linewidth]{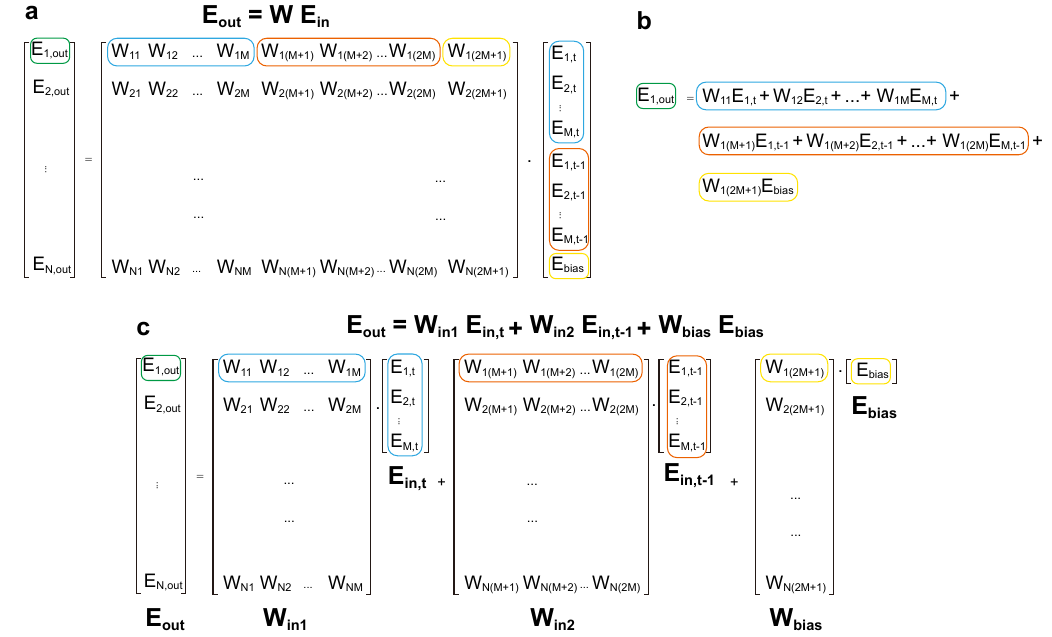}
  } 
  \caption{\noindent\textbf{Illustration of optical NGRC formulation in matrix representation.} \textbf{a} In our optical system, the input field and output field are linearly connected by a transmission matrix. \textbf{b} Every output element (the first is shown as an example) can be represented as the dot product of a row of the transmission matrix and the input field vector. The input elements include the current and previous states as well as the bias. \textbf{c} Similar to breaking out the dot product into sub groups, we can rewrite the matrix vector multiplication the way as presented in the main text.}
 \label{FigureS3}
\end{figure*} 

To illustrate how one can rewrite the optical random projection as Eq. (3) in the main text, we refer to Supplementary Fig. \ref{FigureS3}. As shown in this figure, an output mode is the dot product of a row of the transmission matrix and the input field vector. As such, we can break out the dot product into sub groups containing separately the current input, the previous input and the bias. Similarly, in the matrix form, the computation can also be divided into sub blocks, such that the input matrices $\boldsymbol{W}_{in1}$ and $\boldsymbol{W}_{in2}$ are related to $\boldsymbol{u}_{t}$ and $\boldsymbol{u}_{t-1}$ separately. 

Our optical NGRC scheme can immediately inspire a wide array of physical NGRC based on other various physical substrates. Given a physical reservoir system that performs a nonlinear transformation on its input, say $\boldsymbol{y}=f(\boldsymbol{x})$, we can stimulate the system with time-delayed inputs using our recipe illustrated above, and record the output as reservoir features to define a physical NGRC. More specifically, the physical NGRC can be formulated as $\boldsymbol{r_{t+1}}=f(\boldsymbol{u_t, u_{t-1}})$ (here only two delayed inputs are shown as an example). Through a similar decomposition analysis shown above, one can achieve:
\begin{equation}
\label{eq2}
    \boldsymbol{r}_{t+1} \approx \boldsymbol{M}_s \cdot [1, \underbrace{\boldsymbol{u}_t^T,\boldsymbol{u}_{t-1}^T}_{\mathrm{Linear \thinspace terms}},
    \underbrace{\mathbb{U}(\boldsymbol{u}_{t} \otimes \boldsymbol{u}_{t}), \mathbb{U}(\boldsymbol{u}_{t-1} \otimes \boldsymbol{u}_{t-1}), \mathbb{U}(\boldsymbol{u}_{t} \otimes \boldsymbol{u}_{t-1})}_{\mathrm{Quadratic \thinspace terms}}, ...]^T,
\end{equation}
where $\boldsymbol{M_s}$ is specified by the physical system, and the remaining symbols are consistent with the notation defined in the main text, e.g., $\mathbb{U}$ denotes an operation to collect all unique monomials from the matrix vectorization of the outer product of two vectors. Generalizing our optical NGRC model to a broader regime of physical NGRCs opens up many intriguing directions and holds the potential to enhance the performance of conventional physical RC across diverse scenarios.

\begin{figure*}[ht]
  \renewcommand{\figurename}{Supplementary Figure}
  \centering{
  \includegraphics[width = 1.0\linewidth]{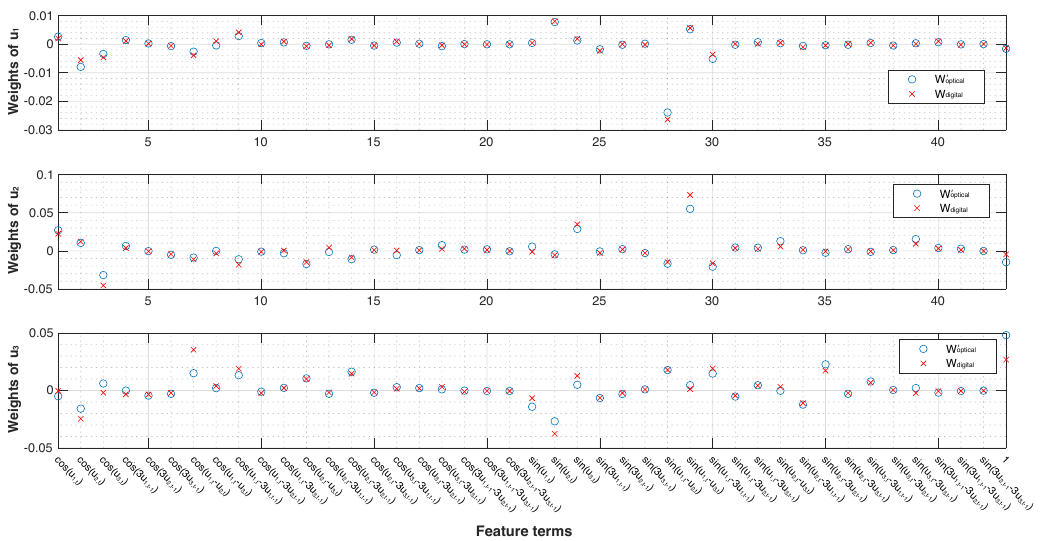}
  } 
  \caption{\noindent\textbf{Comparison of the effective readout matrices of optical NGRC and digital NGRC in the same feature basis.} }
 \label{FigureS4}
\end{figure*} 

At the last part of this note, we quantitatively show that why our optical system is equivalent to NGRC implementation. To this end,we perform additional simulation to investigate the underlying feature extraction process of the optical setup. 
We use the Lorenz63 autonomous forecasting as an example. 
The goal is to compare the weights of the readout matrix obtained in the optical approach with that of the digital NGRC computed in the same feature basis. For this specific task, we encode 7 macropixels to the SLM, including 3 for the current input ($u_{1,t}$, $u_{2,t}$, $u_{3,t}$), 3 for the previous input ($\eta u_{1,t-1}$, $\eta u_{2,t-1}$, $\eta u_{3,t-1}$) and a bias $b$. In the simulation, we use $\eta= 3$ and $b=0$ as an example and the inputs are normalized to $[0,\pi]$, without loss of generality. We also consider the phase encoding ($x \rightarrow{\mathrm{exp}(ix)}$) of the SLM and intensity detection ($x \rightarrow |x|^2$) of the system. While the reservoir size may be large, the optically generated features can be described by the weighted sums of in total 43 independent feature terms, shown in the axis of Supplementary Fig. \ref{FigureS4}. These include 21 cosine terms and 21 sine terms as well as a bias term. For optical NGRC, we first compute the readout matrix $\boldsymbol{W}_{optical}$ and then project them in these 43 feature terms to obtain a new readout matrix $\boldsymbol{W}^{\prime}_{optical}$ . In this way, we can fairly compare the optical NGRC with digital NGRC that is built upon these feature terms directly, since they use the same feature basis. We then train a linear readout layer $\boldsymbol{W}_{digital}$ using digital NGRC. As clearly shown in Supplementary Fig. \ref{FigureS4}, the close matching between $\boldsymbol{W}_{optical}^{'}$ and $\boldsymbol{W}_{digital}$ confirms that the optical NGRC operates in the same way as digital NGRC.

\section*{Supplementary Note 2. Experimental setup details}
\begin{figure*}[!ht]
  \renewcommand{\figurename}{Supplementary Figure}
  \centering{
  \includegraphics[width = 0.5\linewidth]{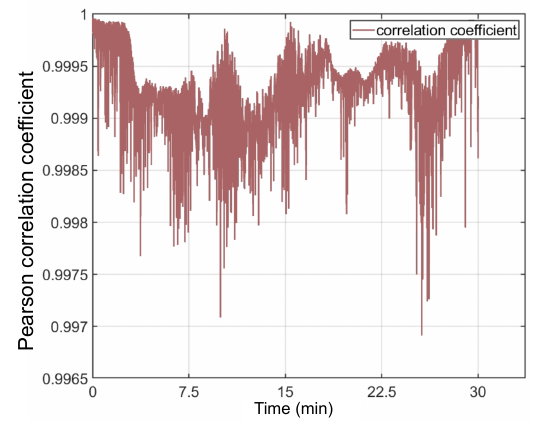}
  } 
  \caption{\noindent\textbf{Experimental system stability.} We calculate the correlations between the initial speckle intensity pattern frame and subsequent frames (reservoir features), all generated using the same phase mask in the SLM.}
 \label{FigureS5}
\end{figure*} 

\begin{figure*}[!ht]
  \renewcommand{\figurename}{Supplementary Figure}
  \centering{
  \includegraphics[width = 0.5\linewidth]{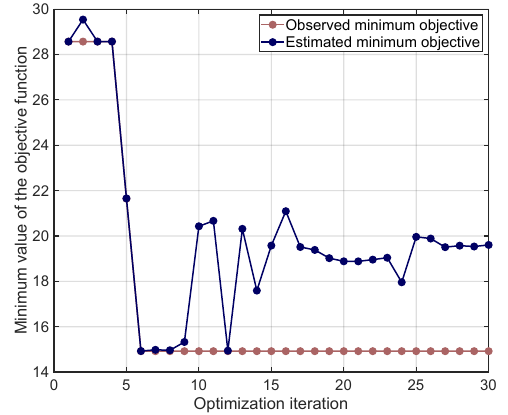}
  } 
  \caption{\noindent\textbf{Bayesian optimization log during the short-term prediction of KS time series experiments.} The objective function is defined as the accumulated error over a certain period of time in the prediction test. The red curve corresponds to the experimentally realized minimum objective function value (up to the number of iterations), while the blue curve indicates what the Bayesian model predicts or estimates, based on the previous experimental realizations.}
 \label{FigureS6}
\end{figure*} 
Here we describe additional experimental details to complement the Methods section. The schematic diagram of the experimental setup is shown in Supplementary Fig. \ref{FigureS1}a and a representative TM is shown in Supplementary Fig. \ref{FigureS1}b. To ensure the accurate phase encoding of input data into optical signals, we calibrate the SLM by updating its lookup table at the optical wavelength of 635 nm. The stability of the optical setup is crucial for the predictions of chaotic time series in optical NGRC. To reduce the experimental noise, we optimize the setup in several aspects. Firstly, we utilize an air conditioner in the lab and a shielding cage enclosing the whole setup to ensure a stable ambient environment. We use only a small central region of the full screen of the SLM to reduce the influence of mechanical vibrations associated with the SLM. Additionally, we place another shielding cage covering both the SLM driver and flex cable to prevent turbulent air flows. A long tube is mounted in front of the camera to block ambient light. To further reduce the influence of noise, for a given phase mask, we repeat the acquisition 4 times and average the results to derive the actual reservoir state. The whole optical system operates at a frame rate of 40 Hz, therefore we can collect 10 reservoir state vectors per second. Taken together, the system is stable enough for consistent computations as an optical reservoir. To quantitatively evaluate the noise level, we collect multiple speckle images with the same input, and then calculate the ratio between the standard deviation of the noise and mean value of the signal from these images, which is experimentally measured to be around $0.0106$. As shown in Supplementary Fig. \ref{FigureS5}, we also characterize the optical system stability by measuring the speckle correlation, which remains greater than 0.995 over a duration of 30 minutes. After optimizing the setup, we use online Bayesian optimization to search for experimental hyperparameters with optical hardware in the loop. An example of the Bayesian optimization process during the experiment is illustrated in Supplementary Fig. \ref{FigureS6}.

\section*{Supplementary Note 3. Simulation comparison of optical NGRC and optical conventional RC based on scattering media}
In the main text, we present the experimental results to show optical NGRC outperforms optical conventional RC based on scattering media in many aspects. Here we conduct additional numerical simulations to further compare these two architectures. We note that the best forecasting performance of the KS time series in simulation (without noise) based on optical conventional RC using scattering media is approximately 4 Lyapunov times, as reported in the Appendix of ref. \cite{rafayelyan2020large}. By using optical NGRC in simulation, we achieve a forecasting capability up to 6 Lyapunov times with a significantly smaller reservoir and reduced training data. To quantitatively show their performance difference, we conduct more simulations as illustrated in Supplementary Fig. \ref{FigureS7}. At a specific reservoir size, we use the same training length (10,000 time steps) for both architectures to ensure a fair comparison between them, and repeat the simulation 25 times with different random matrices. 
Note that the warm-up period in the optical conventional RC is not considered into the training length, which is a bit unfair for the optical NGRC. We use the normalized root mean square error (NRMSE) over the first 2 Lyapunov times of the prediction as the metrics for comparison. As clearly shown in Supplementary Fig. \ref{FigureS7}, optical NGRC consistently achieves better performance (lower error) at all reservoir sizes ranging from 500 to 2,500, and also seems to show a smaller standard deviation than the optical conventional RC based on scattering media. 
\begin{figure*}[!htb]
  \renewcommand{\figurename}{Supplementary Figure}
  \centering{
  \includegraphics[width = 0.5\linewidth]{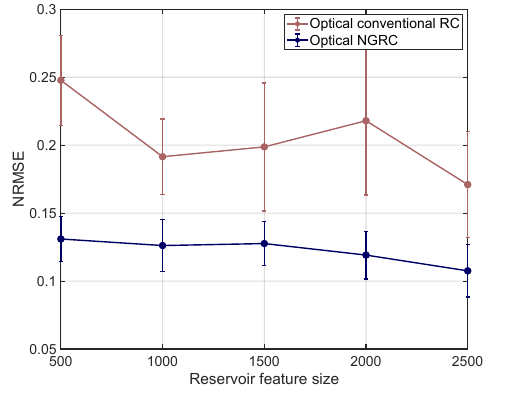}
  } 
  \caption{\noindent\textbf{ Performance comparison of optical NGRC and optical conventional RC based on scattering media in simulation.} At each reservoir size, 25 different realizations are preformed with the same training length of 10,000 time steps. The error bar represents the range of one standard deviation.}
 \label{FigureS7}
\end{figure*} 

\section*{Supplementary Note 4. The impact of device quantization and noise on optical NGRC}
\begin{figure*}[!ht]
  \renewcommand{\figurename}{Supplementary Figure}
  \centering{
  \includegraphics[width = 0.7\linewidth]{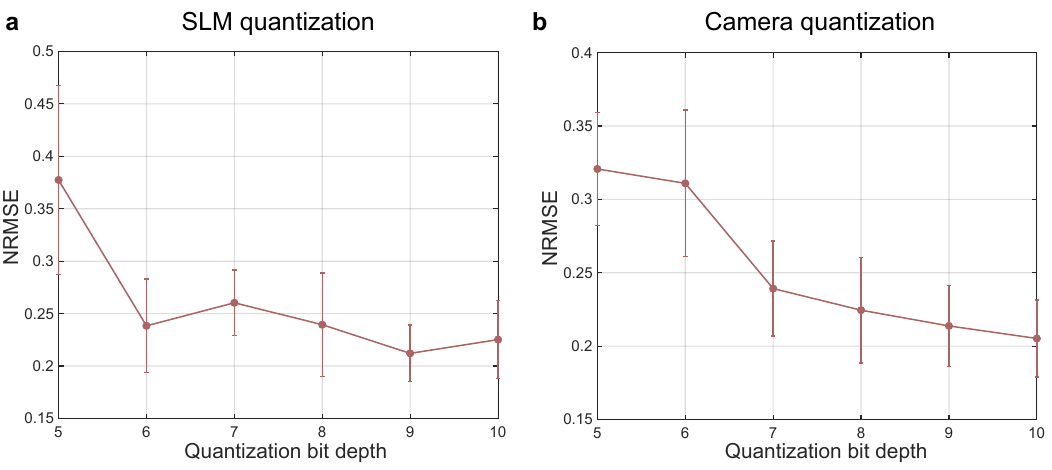}
  } 
  \caption{\noindent\textbf{Simulation of optical NGRC forecasting errors based on different quantization bit depths of devices.} \textbf{a} NRMSE versus quantization bit depth of SLM used for data encoding. \textbf{b} NRMSE versus quantization bit depth of camera used for reservoir feature measurement. At each quantization bit depth, 25 different realizations are preformed with a reservoir size of 2,500 and a training length of 10,000 time steps. The error bar represents the range of one standard deviation. }
 \label{FigureS8}
\end{figure*} 
Here we study the quantization effect of the SLM and camera devices on the optical NGRC performance. Due to the  hardware constraint, quantization leads to data being represented with finite precision, thereby introducing errors into the encoding and detection processes throughout each iteration of the training and test phases. To quantify this effect, similar to Supplementary Note 3, we use the short-term prediction of the KS time series as the target task and calculate the NRMSE over the first 2 Lyapunov times in the prediction phase. With a reservoir size of 2,500 and a training length of 10,000 time steps, we examine the impact of quantization of two devices separately (by treating the other device free from quantization errors). As shown in Supplementary Fig. \ref{FigureS8}, it is expected that optical NGRC predicts better with lower errors with increased bit depths. As stated in the main text, our setup currently employs an effective bit depth of 7 bits for the SLM and 8 bits for the camera. This quantization level in the experiment is sufficient for the proof-of-concept demonstrations (both short-term and long-term forecasting and NGRC observer) in this work. 

\begin{figure*}[!ht]
  \renewcommand{\figurename}{Supplementary Figure}
  \centering{
  \includegraphics[width = 0.5\linewidth]{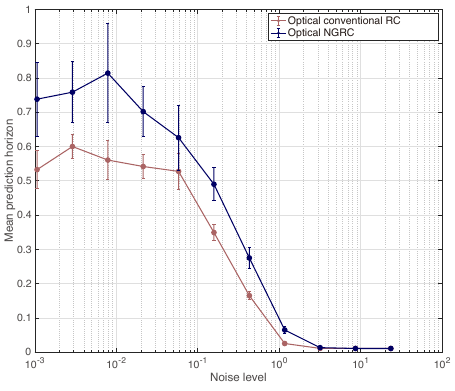}
  } 
  \caption{\noindent\textbf{Performance comparison of optical NGRC and optical conventional RC based on scattering media with added noise to the reservoir features in simulation.} The noise level represents the ratio between the standard deviation of the added Gaussian noise and the mean value of the reservoir signal. The mean prediction horizon denotes the prediction length when NRMSE reaches a threshold of 0.2. The error bar indicates the range of one standard deviation from 40 repetitive realizations. }
 \label{FigureS9}
\end{figure*} 

In addition, we also explore how the noise added to the reservoir features could impact the performance. Similar to Supplementary Fig. \ref{FigureS7}, here we compare the optical NGRC with conventional optical RC based on scattering media. As shown in Supplementary Fig. \ref{FigureS9}, we use the mean prediction horizon as the metric, characterized by the average prediction length when NRMSE reaches a threshold of 0.2. The reason why we use the mean prediction horizon here instead of NRMSE is that this metric is relatively easier to visualize, thanks to its bounded variance, which makes it more suitable for assessing the impact of noise. It can be seen that for both RC systems, an appropriate level of noise enhances forecasting performance due to noise regularization \cite{estebanez2019constructive}, but the prediction length decreases significantly when the noise is too strong. Under almost all noise levels of interest, the optical NGRC outperforms (predicts longer than) conventional RC based on scattering media.

\section*{Supplementary Note 5. Optical computation analysis}
Here we analyze the computational operations performed by the experimental setup and investigate its potential scaling properties. In this work, we exploit multiple light scattering to compute the reservoir features efficiently. As introduced in Supplementary Note 1, we perform a random projection optically described by a complex matrix $\boldsymbol{W} \in \mathbb{C}^{N\times M}$ where $M$ and $N$ are the number of input and output modes, respectively.
We can break down each complex computational operation into its constituent real operations, wherein a complex multiplication is decomposed into 4 real multiplications and 2 real additions, and a complex addition entails 2 real additions \cite{zhou2021large}. We omit the element-wise computations related to encoding and detection nonlinearity since they correspond to a comparatively small number of operations. Given that the optical reservoir feature extraction encompasses $NM$ complex multiplications and $N(M-1)$ complex additions, the equivalent total is $6NM+2N(M-1)=8NM-2N$ real operations. For example, in the KS system forecasting experiments where $M=64\times2=128$ (two time steps) and $N=2,500$, the setup achieves approximately $0.1$ giga floating point operations per second (GFLOPS) at a system frame rate of 40 Hz. Regarding the power consumption, in our experiment, the laser output power is approximately 2.5 mW, while the power usage is around 20 W for the SLM, 2.5 W for the camera, and 50 W for the control desktop computer, cumulating in a total power usage of 72.5 W. As a result, we estimate the computation energy efficiency as $\eta=102,200,000/72.5025\approx1.41$ MegaOp/J (or equivalently $0.71$ $\mu$J/Op).

Improvements in energy efficiency can be achieved by  reducing the power consumption of each component and/or increasing the overall frame rate of the system. This can be achieved by replacing the currently employed devices with more advanced alternatives. Currently we use only a small central region of the full screen of the SLM, therefore an SLM with less pixels but faster frame rate is preferred for our demonstration. For instance, using an one-dimensional SLM based on grating light valve can experimentally achieve 350 kHz modulation speed \cite{tzang2019wavefront} and an electro-optic SLM can reach a GHz frame rate \cite{trajtenberg2024lnos} and potentially increase the system's processing speed by orders of magnitude. Likewise, the system's overall frame rate also depends on the speed of the camera, such that a faster detection device (e.g., a silicon photonic detector array with sub-GHz bandwidth \cite{rogers2021universal} and a single-photon avalanche diode camera with sub hundred ps timing resolution \cite{morimoto2020megapixel}) could enhance the computational performance as well. In addition, the desktop computer used for managing the digital backend and communicating multiple devices can be replaced by a more energy-efficient electronic device, potentially with a power consumption below 10 W \cite{li2022fpga}. With these aforementioned hardware, we could envision a speed up of our system to at least kHz. Collectively, the total power consumption and the overall frame rate can be within approximately $10$ (SLM from ref. \cite{trajtenberg2024lnos}) $+10$ (FPGA from ref.\cite{li2022fpga}) $+2$ (Camera) $+0.0025$ (Laser) $\approx 22$ W and reach 4 kHz (100 times higher than the current system), resulting in an energy efficiency of 2.2 nJ/Op. Note that the optical computation scale is determined by the pixel numbers of the SLM and camera used in the experiment. Luckily, the current technologies of SLMs and cameras support megapixels, which are already sufficient to show the optical computing advantage \cite{rafayelyan2020large}.
Moreover, even the aforementioned hardware optimization support million pixels, therefore the scaling can still be optimistically expected. Nevertheless, we cannot estimate the cost of the envisioned system as some of the devices are still under development in the laboratory. While the overall signal-to-noise ratio (SNR) of the system does not necessarily degrade when towards larger dimension and faster clock rate, the SNR of the system depends on the individual components used in the experiment. 
As a side note, we want to point out that replacing the SLM with a fast digital micromirror device (DMD) is not a viable solution here, as the encoding binarization could introduce large errors, thereby compromising performance particularly for challenging autonomous forecasting tasks. However, it is indeed interesting to explore the use of the fast and efficient system based on DMDs for other machine learning tasks rather than autonomous time-series forecasting, such as classification, based on optical NGRC in future studies.

Although we may not be able to compete with advanced commercial graphics processing unit in the current setting, such as NVIDIA V100 TENSOR CORE that achieves $0.27$ TeraOp/J ($3.7$ pJ/Op) \cite{khairy2020tpu}, the optical NGRC features favourable scaling properties. In our system,  the optical computation time and memory requirements almost do not scale with the reservoir dimension $N$, i.e. $O(1)$. In practice, they scale linearly with the reservoir size, i.e. $O(N)$, considering the digital-to-analog and analog-to-digital conversions in the system. On the contrary, digital computers based on the von Neumann architecture exhibits quadratic scaling, i.e. $O(N^2)$, for matrix-vector multiplications. Therefore, we can foresee the optical setup will surpass digital computers in speed and efficiency beyond a certain data dimension threshold. This potential has been clearly evidenced by several previous experimental studies, where the benefits of optical computing emerge as $N$ approaches the order of $10^4$ \cite{ohana2020kernel,rafayelyan2020large,pierangeli2021scalable}. The current technologies of commercial SLMs and cameras are both at the megapixel ($10^6$ pixels) level. As such, we can encode a large amount of data in different sizes with a megapixel SLM and compute rich features with a megapixel camera. This can correspond to a $10^6 \times 10^6$ complex transmission matrix in our computing system, relating the input field at the SLM plane and output field at the camera plane. Computing the matrix vector multiplication at such a scale will be exhausted for digital computers in terms of both speed and memory costs.

The computational cost of a digital NGRC is quite different from optical NGRC. 
Firstly, considering an input data vector $\boldsymbol{u_t}\in \mathbb{R}^M$, if we build the NGRC feature vector $\boldsymbol{r_t}$ from $K$ time steps and up to a polynomial order of $H$, the total number of feature terms is calculated as $N^{\prime}=\sum_{i=0}^{H} \frac{(MK+i-1)!}{i!(MK-1)!}$. More precisely, for the low-dimensional Lorenz63 forecasting task in the main text, if we use $M=3$, $K=2$ and $H=2$ as in ref. \cite{gauthier2021next}, the reservoir size will be $N^{\prime}=28$. Similarly, for predicting the KS time series, the reservoir size is calculated to be $N^{\prime}=8,385$ for $M=64$, $K=2$ and $H=2$. 
As a result, the size of the reservoir feature $N^{\prime}$ increases polynomially with the data dimension $M$, and the ridge regression at a large reservoir size during training will become inefficient due to matrix inversion with complexity of $O({N^{\prime}}^3)$. And the total computational cost for each inference step can be estimated as $O(N^{\prime}M)$. Heuristically speaking, the computational cost of digital NGRC increases polynomially when scaling to a large dimension since $N^{\prime}$ scales polynomially with $M$, if without prior knowledge of the data system to perform dimension reduction.
A possible solution has been proposed to tackle this challenge, which employs a number of parallel NGRC to learn the large-scale KS time series \cite{barbosa2022learning}. In comparison, although the total computational complexity of optical NGRC is similarly $O(NM)$ dominated by the digital readout, $N$ could be much smaller than that of the digital NGRC towards the large dimension. 
For instance, for the same KS prediction studied in this work, the digital NGRC would require the reservoir size to be $N^{\prime} = 8,385$, while we only need $N=2,500$ in optical NGRC. 
Consequently, the scalability of optical NGRC in accommodating large input dimensions can be quite different from that of digital NGRC systems. As a side note, the polynomial feature terms in optical NGRC are generated naturally within each speckle feature on the camera, eliminating the need for the manual determination of the potential polynomial orders to be used. Instead, we simply train a linear readout layer to retrieve the most relevant feature terms. In optical NGRC, we can flexibly adapt the reservoir size depending on the difficulty of the task and the target performance. Yet, our optical NGRC approach requires a considerably larger reservoir size than the digital NGRC for small-scale datasets like the Lorenz63 system. This implies that we would need longer training length than the digital NGRC to obtain the readout matrix, as indicated in Supplementary Table S1. To tackle this issue, it will be beneficial to explore feature selection \cite{cai2018feature} on the experimentally generated reservoir feature in future studies.

Lastly, we discuss how the limitation of the feedback between the camera and the SLM impacts our current system. For the training phase, one advantage of our optical NGRC scheme compared to conventional optical RC \cite{rafayelyan2020large} is that we do not require feedback between the camera and SLM. That is, we can obtain all the reservoir states in training without knowing previous reservoir states, thus alleviating the feedback bottleneck. 
In contrast, for conventional optical RC, one has to wait for $\boldsymbol{r}_{t}$ to be collected and processed, and then obtain $\boldsymbol{r}_{t+1}$ afterwards. This can be limited by the speed of communication between SLM and camera. 
For prediction (inference), such feedback bottleneck impacts differently for autonomous and non-autonomous prediction tasks. 
If we are dealing with autonomous forecasting tasks (Figs. 2 and 3 of the main text), we need to calculate the output $\boldsymbol{o}_{t+1}$ digitally with a trained readout matrix for each time step, based on the optically-generated reservoir state $\boldsymbol{r}_{t+1}$. 
And then we take this output as the input $\boldsymbol{u}_{t+1}$ for the next time step. As such, the feedback indeed impacts the autonomous forecasting tasks and slows down the overall frame rate of the system. But note that this digital and communication overhead scales linearly with the reservoir state sizes, as demonstrated in, e.g., Fig. 6 of ref. \cite{rafayelyan2020large}. For large-scale reservoir size, it is still possible to achieve the optical advantage. 
If we are dealing non-autonomous tasks, for instance the observer task in Fig. 4 of the main text, where we do not have to wait for the previous output to calculate the next output, we do not have this optical-electronic-optical feedback and are not constrained by such digital bottleneck just as the training phase, as opposed to optical conventional RC.

\hfill \break
\section*{Supplementary Algorithms:}

\begin{algorithm}[H]
    \SetAlgoLined
    \caption{Optical NGRC for forecasting dynamical systems}
    \KwResult{Predictions $\{ \boldsymbol{\hat{o}}_t \} \in \mathbb{R}^{T_{test}\times M}$}
    \KwIn{A training set $\{\boldsymbol{u}_t \} \in \mathbb{R}^{T_{train}\times M} $ } 
    \textbf{Training:} Prepare training ground truth $\{ \boldsymbol{o}_t \} \in \mathbb{R}^{(T_{train}-2)\times M}$ based on $\boldsymbol{o}_t = \boldsymbol{u}_{t+2}$\;
    \For{$t=2,3,...,T_{train}$}{
    Compute the SLM phase mask based on $[\boldsymbol{u}_t,\boldsymbol{u}_{t-1},b]^T$\;
    Run the optical experimental setup to retrieve the reservoir state $\boldsymbol{r}_{t+1} \in \mathbb{R}^N$\;
    }
    Compute the output layer $\boldsymbol{W}_{out} \in \mathbb{R}^{M\times N}$ by minimizing $\Vert   \boldsymbol{W}_{out} \{ \boldsymbol{r}_t \}- \{\boldsymbol{o}_t\} \Vert_2^2 + \beta\Vert \boldsymbol{W}_{out}\Vert_2^2$\;
\textbf{Prediction:} Initialize a prediction starting point by specifying $\boldsymbol{u}_2$ and $\boldsymbol{u}_1$ as the last two time steps of the training set\;
    \For{$t=2,3,...,T_{test}+1$}{
    Compute the SLM phase mask based on $[\boldsymbol{u}_t,\boldsymbol{u}_{t-1},b]^T$\;
    Run the optical experimental setup to retrieve the reservoir state $\boldsymbol{r}_{t+1} \in \mathbb{R}^N$\;
    Compute the prediction based on $\boldsymbol{\hat{o}}_{t+1} =\boldsymbol{W}_{out} \boldsymbol{r}_{t+1}$\;
    Assign $\boldsymbol{\hat{o}}_{t+1}$ to $\boldsymbol{u}_{t+1} $\;
    }
Return the predictions $\{ \boldsymbol{\hat{o}}_t \}$\
\end{algorithm}

\vspace{1cm}

\begin{algorithm}[H]
    \SetAlgoLined
    \caption{Optical NGRC for deducing unmeasured variables of dynamical systems}
    \KwResult{Predictions $\{ \boldsymbol{\hat{o}}_t \} \in \mathbb{R}^{T_{test}\times Q}$}
    \KwIn{A training input set $\{\boldsymbol{u}_t \} \in \mathbb{R}^{T_{train}\times P} $ with training ground truth $\{ \boldsymbol{o}_t \} \in \mathbb{R}^{T_{train}\times Q}$, a test input set $\{\boldsymbol{v}_t \} \in \mathbb{R}^{T_{test}\times P} $ } 
    \textbf{Training:} Determine the number of input time steps as 5 spaced with a stride of 5\;
    \For{$t=21,22,...,T_{train}$}{
    Compute the SLM phase mask based on $[\boldsymbol{u}_t,\boldsymbol{u}_{t-5},\boldsymbol{u}_{t-10},\boldsymbol{u}_{t-15},\boldsymbol{u}_{t-20},b]^T$\;
    Run the optical experimental setup to retrieve the reservoir state $\boldsymbol{r}_t \in \mathbb{R}^N$\;
    }
    Compute the output layer $\boldsymbol{W}_{out} \in \mathbb{R}^{Q\times N}$ by minimizing $\Vert   \boldsymbol{W}_{out} \{ \boldsymbol{r}_t \}- \{\boldsymbol{o}_t\} \Vert_2^2 + \beta\Vert \boldsymbol{W}_{out}\Vert_2^2$\;
\textbf{Prediction:} Initialize the test starting point by drawing 5 time steps from the tail of the training input set as $[\boldsymbol{v}_{21},\boldsymbol{v}_{16},\boldsymbol{v}_{11},\boldsymbol{v}_{6},\boldsymbol{v}_{1}]$\;
    \For{$t=21,22,...,T_{test}+20$}{
    Compute the SLM phase mask based on $[\boldsymbol{v}_t,\boldsymbol{v}_{t-5},\boldsymbol{v}_{t-10},\boldsymbol{v}_{t-15},\boldsymbol{v}_{t-20},b]^T$\;
    Run the optical experimental setup to retrieve the reservoir state $\boldsymbol{r}_t \in \mathbb{R}^N$\;
    }
Compute the prediction based on $\{\boldsymbol{\hat{o}}_{t} \}=\boldsymbol{W}_{out} \{\boldsymbol{r}_{t} \}$ \;
Return the predictions $\{ \boldsymbol{\hat{o}}_t \}$\
\end{algorithm}

\vspace{3cm}

\hfill \break
\section*{Supplementary Tables:}
\begin{table}[hbt!]
\renewcommand{\tablename}{Supplementary Table}
\begin{threeparttable}
\caption{Comparison of optical NGRC, optical conventional RC based on scattering media and digital NGRC 
\label{Table combined}
}
% \begin{center}
\scriptsize
\centering
\begin{tabular}{ m{3.6cm}<{\centering} m{3.6cm}<{\centering} m{3.6cm}<{\centering} m{3.6cm}<{\centering}  }
\hline
\textbf{Metrics} & \textbf{Digital NGRC \cite{gauthier2021next}} & \textbf{Optical conventional RC \cite{rafayelyan2020large}} & \textbf{Optical NGRC (this work)}  \\
\hline
Training length & Short\tnote{a} & Long (90,500 time steps for KS prediction) & Moderate (6,000 time steps for KS prediction)\tnote{b} \\
Warm-up before training & Very short (typically 2) & Long (typically $10^2 \sim 10^5$) & Very short (typically 2) \\
Number of hyperparameters & 2\tnote{c} & 6\tnote{d} & 3\tnote{e} \\
Forecasting performance & Around 6 time units for Lorenz63\tnote{f} & Around 4 Lyapunov times for KS system & Around 2.5 Lyapunov times for KS sysem \\
Reservoir feature size & 28 for Lorenz63 system & 2,500 for KS system & 10,000 for KS system \\
Model interpretability\tnote{g} & Interpretable & Uninterpretable & Interpretable \\
Physical openness\tnote{h} & Incompatible & Compatible & Compatible \\

\hline
\end{tabular}
\begin{tablenotes}
\footnotesize
\item[a] The training length of digital NGRC is shorter than conventional digital RC for low-dimensional chaotic time series processing  \cite{pathak2017using,gauthier2021next}, such as in Lorenz63 forecasting.
\item[b] The training length of optical NGRC is usually longer than that of digital NGRC due to the larger reservoir size and more readout parameters are used. 
\item[c] The hyperparameters include the order of polynomials and ridge regularization parameter.
\item[d] The hyperparameters include the encoding scaling factors $s_{in}$ and $s_{res}$, the encoding micropixel sizes $p_{in}$ and $p_{res}$, the leaking rate, and the ridge regularization parameter \cite{rafayelyan2020large}. 
\item[e] The hyperparameters include the relative weight $\eta$ between two inputs, the bias $b$, and the ridge regularization parameter. 
\item[f] The forecasting performance is comparable to the results previously achieved in ref. \cite{pathak2017using}.
\item[g] Interpreting conventional RC is challenging \cite{nakajima2021reservoir}, whereas both digital and optical NGRC are considered interpretable, as they exploit features from time-delayed inputs for applications.
\item[h] Physical openness refers to the compatibility and adaptability of the RC scheme to (other) physical hardware.

\end{tablenotes}

% \end{center}
\end{threeparttable}
\end{table}

\begin{table}[hbt!]
\begin{threeparttable}
\renewcommand{\tablename}{Supplementary Table}
\caption{Summary of data encoding and processing parameters used in the experiments  
\label{Table combined2}
}
% \begin{center}
\scriptsize\centering
\begin{tabular}{ m{3.6cm}<{\centering} m{2.9cm}<{\centering} m{2.8cm}<{\centering} m{2.8cm}<{\centering} m{2.8cm}<{\centering}}
\hline
\textbf{Parameters} & \textbf{Lorenz63 forecasting}\tnote{a} & \textbf{KS forecasting}\tnote{b} & \textbf{Lorenz63 observer}\tnote{c} & \textbf{KS observer}\tnote{d}  \\
\hline
Input bias ($b$) & 1.6 & 1.1 & 1.5 & 1.5 \\
Relative weight ($\eta$) & 7.5$\times 10^{-1}$ & 9.7$\times 10^{-1}$  & 1.0 & 1.0 \\
Number of time steps & 2 & 2 & 5 & 5 \\
Encoding macropixel & 28$\times$28 & 7$\times$7 & 28$\times$28 & 21$\times$21 \\
Speckle grain size (pixels) & 7 & 7 & 7 & 7 \\
Reservoir size & 2,000 & 2,500 & 2,000 & 2,500 \\
Training length & 4,000 & 6,000 & 4,00 & 10,000 \\
 Time interval ($\Delta t$) & 0.025 & 0.25 & 0.025 & 0.25 \\
Regularization parameter ($\beta$) & 1.5$\times 10^{-1}$ & 5.6$\times 10^{-1}$ & 4.3$\times 10^{-5}$ & 3.4$\times 10^{-1}$  \\

\hline
\end{tabular}
\begin{tablenotes}
\footnotesize
\item[a] The parameters used in short-term forecasting of Lorenz attractor in Fig. 2c.
\item[b] The parameters used in short-term forecasting of KS system in Fig. 3a.
\item[c] The parameters used in Lorenz63 observer in Fig. 4b.
\item[d] The parameters used in KS observer in Fig. 4c.

\end{tablenotes}

% \end{center}
\end{threeparttable}
\end{table}

\begin{table}[hbt!]
\begin{threeparttable}
\caption{Performance comparison with previous works on Lorenz63 and KS time-series prediction
\label{Table S3}
}
\scriptsize\centering
\begin{tabular}{ m{7.3cm}<{\centering} m{3.8cm}<{\centering} m{3.8cm}<{\centering}  }
\hline
\textbf{Reference} & \textbf{Lorenz63 forecasting} & \textbf{KS forecasting} \\
\hline
Gauthier et al. (Numerical) \cite{gauthier2021next} & $\sim$ 6 time units & Not available \\
Pathak et al. (Numerical) \cite{pathak2018model} & Not available & $\sim$ 6 Lyapunov times\\
Pathak et al. (Numerical) \cite{pathak2017using} & $\sim$ 6 time units & $\sim$ 5 Lyapunov times \\
Vlachas et al. (Numerical) \cite{vlachas2020backpropagation} & Not available & $\sim$ 4 Lyapunov times \\
Lu et al. (Numerical) \cite{lu2018attractor} & $\sim$ 7 time units & Not available \\
Wikner et al. (Numerical) \cite{wikner2021using} & $\sim$ 4 time units & $\sim$ 2.5 Lyapunov times \\
Jiang et al. (Numerical) \cite{jiang2019model} & Not available & $\sim$ 6 Lyapunov times \\
Dong et al. (Numerical) \cite{dong2020reservoir} & Not available & $\sim$ 5 Lyapunov times \\
Antonik et al. (Experimental) \cite{PhysRevApplied.7.054014} & $\sim$ 2 time units & Not available \\
Rafayelyan et al. (Experimental) \cite{rafayelyan2020large}
& Not available & $\sim$ 2.5 Lyapunov times \\
\textbf{This work (Experimental)} & $\sim$ 4.5 time units & $\sim$ 4 Lyapunov times \\

\hline
\end{tabular}
\end{threeparttable}
\end{table}

\renewcommand{\bibpreamble}{
$^\ast$These authors contributed equally to this work.\\
$^\dagger${Corresponding authors: \textcolor{magenta}{jianqi.hu@epfl.ch}, \textcolor{magenta}{qiangliu@tsinghua.edu.cn}, \textcolor{magenta}{sylvain.gigan@lkb.ens.fr}}\\
}

\bibliographystyle{naturemag}
\bibliography{ref}